\documentclass[review,12pt]{elsarticle}

\usepackage{amssymb}
\usepackage{amsthm}
\usepackage{framed} 
\usepackage{amsmath,color}
\usepackage{mathrsfs}
\usepackage{graphicx}
\usepackage{epstopdf}
\usepackage{float}
\usepackage{caption}
\usepackage{subcaption}
\usepackage{bm}
\usepackage{bbm}
\usepackage{mathrsfs}
\usepackage{cleveref}
\usepackage{soul}
\usepackage{accents}
\usepackage{color,soul} 
\usepackage{color} 
\usepackage{bm}
\usepackage{multirow} 
\biboptions{sort&compress}
\soulregister\citep7 
\soulregister\citet7 
\soulregister\citealp7 
\newsavebox{\measurebox} 
\usepackage{titlesec} 
\usepackage[T1]{fontenc} 
\usepackage{lmodern} 
\input{glyphtounicode}  

\newcommand{\tm}{\textrm} 

\def\onedot{$\mathsurround0pt\ldotp$}
\def\cddot{
  \mathbin{\vcenter{\baselineskip.67ex
    \hbox{\onedot}\hbox{\onedot}}%
  }}%
\def\cdddot#1{
  \mathbin{\vcenter{\baselineskip.67ex
    \hbox{\onedot}\hbox{\onedot}\hbox{\onedot}%
  }}%
}

\journal{Theoretical and Applied Fracture Mechanics}

\makeatletter
\def\@author#1{\g@addto@macro\elsauthors{\normalsize%
    \def\baselinestretch{1}%
    \upshape\authorsep#1\unskip\textsuperscript{%
      \ifx\@fnmark\@empty\else\unskip\sep\@fnmark\let\sep=,\fi
      \ifx\@corref\@empty\else\unskip\sep\@corref\let\sep=,\fi
      }%
    \def\authorsep{\unskip,\space}%
    \global\let\@fnmark\@empty
    \global\let\@corref\@empty  
    \global\let\sep\@empty}%
    \@eadauthor={#1}
}
\makeatother

\begin{document}

\begin{frontmatter}



\title{Applications of phase field fracture in modelling hydrogen assisted failures}


\author{Philip K. Kristensen \fnref{DTU}}

\author{Christian F. Niordson\fnref{DTU}}

\author{Emilio Mart\'{\i}nez-Pa\~neda\corref{cor1}\fnref{IC}}
\ead{e.martinez-paneda@imperial.ac.uk}

\address[DTU]{Department of Mechanical Engineering, Technical University of Denmark, DK-2800 Kgs. Lyngby, Denmark}

\address[IC]{Department of Civil and Environmental Engineering, Imperial College London, London SW7 2AZ, UK}

\cortext[cor1]{Corresponding author.}

\begin{abstract}
The phase field fracture method has emerged as a promising computational tool for modelling a variety of problems including, since recently, hydrogen embrittlement and stress corrosion cracking. In this work, we demonstrate the potential of phase field-based multi-physics models in transforming the engineering assessment and design of structural components in hydrogen-containing environments. First, we present a theoretical and numerical framework coupling deformation, diffusion and fracture, which accounts for inertia effects.Several constitutive choices are considered for the crack density function, including choices with and without an elastic phase in the damage response. The material toughness is defined as a function of the hydrogen content using an atomistically-informed hydrogen degradation law. The model is numerically implemented in 2D and 3D using the finite element method. The resulting computational framework is used to address a number of case studies of particular engineering interest. These are intended to showcase the model capabilities in: (i) capturing complex fracture phenomena, such as dynamic crack branching or void-crack interactions, (ii) simulating standardised tests for critical components, such as bolts, and (iii) enabling simulation-based paradigms such as \emph{Virtual Testing} or \emph{Digital Twins} by coupling model predictions with inspection data of large-scale engineering components. The evolution of defects under in-service conditions can be predicted, up to the ultimate failure. By reproducing the precise geometry of the defects, as opposed to re-characterising them as sharp cracks, phase field modelling enables more realistic and effective structural integrity assessments.
\end{abstract}


\end{frontmatter}



\section{Introduction}
\label{Sec:Introduction}

Hydrogen has been known for decades to notably reduce the toughness, ductility and fatigue life of engineering components \cite{Johnson1875,Gangloff2003,AM2020}. Hydrogen ingress into a metallic sample can happen during its initial forming, during the coating or plating of a protective layer, through exposure to hydrogen or hydrogen-containing molecules in the air, soil or water, or through corrosion processes. Hydrogen absorbed from gaseous or aqueous environments diffuses within the metal and is attracted to areas of high hydrostatic stress, where damage takes place by means of mechanisms that are still being debated \cite{Birnbaum1994,Harris2018,JMPS2019,Shishvan2020}. This so-called hydrogen embrittlement phenomenon is now pervasive across applications in the construction, defence, transport and energy sectors, due to the ubiquity of hydrogen and the higher susceptibility of modern, high-strength alloys \cite{Gangloff2012,Djukic2019}.\\

The use of fracture mechanics-based models for hydrogen-sensitive applications could be a game-changer in preventing catastrophic failures and optimising material performance. For example, reliable modelling of hydrogen assisted fracture could enable a controlled use of high strength alloys, accelerate material certification, and govern inspection planning and fitness-for-service assessment. Yet, the development of models capable of predicting crack initiation and growth as a function of material, loading and environmental variables has not been an easy task. Two main challenges hold back the use of predictive models in engineering assessment. The first one is the physical complexity of the problem at hand. Hydrogen embrittlement is a complicated chemical and micro-mechanical phenomenon that involves multiple hydrogen-metal interactions at several scales. However, several mechanistic models have been proposed that show good agreement with experiments with little or no calibration \cite{Serebrinsky2004,Novak2010,AM2016}. Predictions based on nominal material properties and parameters that can be independently determined are now possible. The second challenge lies in developing a computational framework capable of capturing, in arbitrary geometries and dimensions, the multi-physics elements of the problem and their interaction with the complex cracking phenomena occurring in engineering applications. The phase field fracture method \cite{Bourdin2000} appears to provide a suitable framework for overcoming this obstacle.\\

\begin{figure}[H]
    \centering
    \includegraphics[scale=0.8]{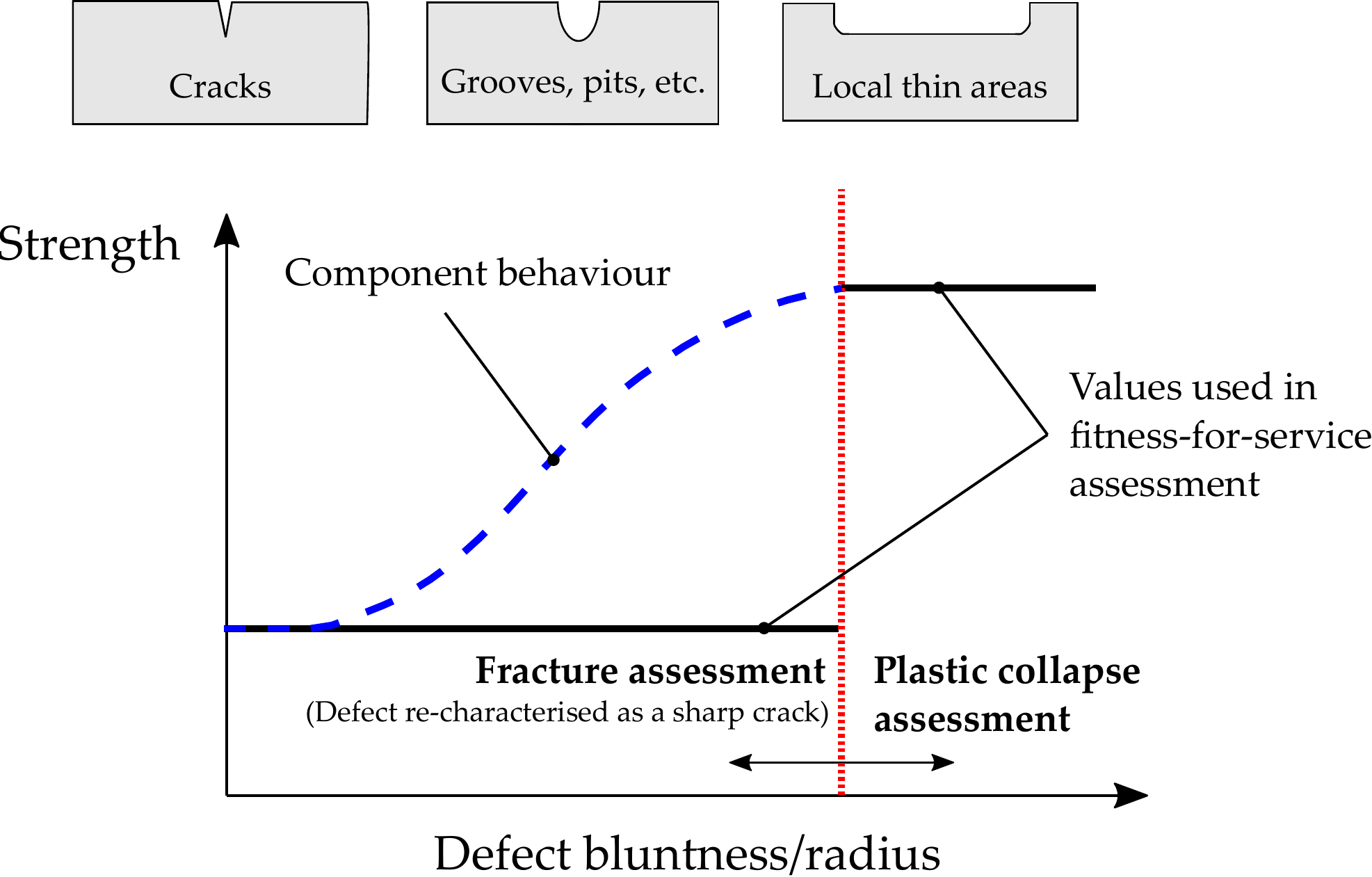}
    \caption{Using phase field to enable defect mechanics-based assessments that can reduce overly conservative fitness-for-service assessment. Schematic of current assessment procedures for a defect of length larger than the transition flaw size; adapted from standardised engineering assessment of burst pressure, see Ref. \cite{Larrosa2018}.}
    \label{fig:DefectMechanics}
\end{figure}

Mart\'{\i}nez-Pa\~neda \textit{et al.} \cite{CMAME2018} have recently extended the phase field fracture method to predict hydrogen assisted failures and the approach has quickly gained popularity as a framework for incorporating various hydrogen embrittlement models \cite{Duda2018,Anand2019,Wu2020b,CS2020,Huang2020,JMPS2020}. The main experimental trends have been captured and advanced fracture features such as crack merging, nucleation from arbitrary sites and branching are predicted without convergence problems or the need for remeshing. However, the vast majority of the analyses are restricted to 2D boundary value problems of mostly academic relevance. In this work, we demonstrate the potential of the phase field fracture method in predicting large scale hydrogen assisted failures of practical engineering interest. This includes, for the first time, (i) the modelling of hydrogen assisted fractures resulting from dynamic loading, where inertia is relevant, and (ii) the consideration of both AT1 \cite{Pham2011} and AT2 \cite{Ambrosio1991,Bourdin2000} constitutive choices for the dissipation function of the phase field fracture method. Overall, the goal is to showcase the capabilities of phase field-based hydrogen assisted cracking formulations in enabling \emph{Virtual Testing} in hydrogen-sensitive applications. To this end, case studies will be addressed involving (i) crack branching in a hydrogen embrittled plate subjected to dynamic loading, (ii) crack-void interactions in a hydrogen-containing 3D bar, (iii) failure of a screw anchor exposed to an aggressive solution, simulating a standardised experiment, and (iv) cracking evolution in a pipeline with internal defects, as measured by in-line inspection. The last example showcases the possibility of combining phase field modelling with inspection data to create (so-called) \emph{Digital Twins} of critical infrastructure, minimising expensive testing and monitoring. In this regard, there is a further motivation for the use of phase field for engineering assessment. Unlike other computational approaches, such as discrete fracture methods \cite{Yu2016a,Diaz2017,EFM2017}, predictions are not restricted to the evolution of sharp cracks but the growth of defects of any arbitrary shape can be simulated, and without any prior knowledge regarding the extent of growth or the growth direction. This opens the possibility of conducting \emph{defect mechanics}-based assessments of notch-like defects, significantly reducing the conservatism associated with re-characterising all detected defects into sharp cracks. The concept is illustrated in Fig. \ref{fig:DefectMechanics} for a given defect length (larger than the transition flaw size); simulating the real defect geometry, as opposed to an equivalent crack, can provide more realistic and sustainable criteria for engineering assessment.

The remainder of this paper is organised as follows. A generalised phase field framework for chemo-mechanical fracture is given in Section \ref{Sec:Theory}, including inertia effects. Details of the finite element implementation are given in Section \ref{Sec:Numerical}. The four case studies described above are presented in Section \ref{Sec:Results}. Finally, the manuscript ends with concluding remarks in Section \ref{Sec:Conclusions}.

\section{A phase field fracture formulation for hydrogen embrittlement}
\label{Sec:Theory}

\subsection{Potential energy of the solid}

Consider a domain $\Omega\in \rm I\!R^n$ $(n \in[1,2,3])$, with outer boundary $\delta\Omega \in \rm I\!R^{n-1}$. The domain contains a deformable solid with displacement field $\bm{u}$ and internal crack surface $\Gamma \in \rm I\!R^{n-1}$. An absorbed species of concentration $c$ might diffuse through the solid and interact with the mechanical behaviour. The variational energy functional for the solid can be postulated as \citep{Francfort1998,Borden2012}: 
\begin{equation}\label{eq:GriffithPotential}
     \Pi = \int_{\Omega}\left\{\frac{1}{2}\rho \dot{\bm{u}}\cdot \dot{\bm{u}} + \psi_s\left(\bm{u},c\right) + \psi_{chem}\left(\bm{u},c\right)  \right\}\,\tm{d}V + \int_{\Gamma}G_c\,\tm{d}S \, .
\end{equation}
Here, the first term of the volume integral represents the kinetic energy density of the solid, with $\dot{\bm{u}}$ denoting the velocity field and $\rho$ the mass density. Also within the volume integral, the second and third terms respectively denote the strain energy density and the chemical energy density related to the transport of solute species. The surface integral represents the fracture energy as proposed by Griffith \cite{Griffith1920}, with $G_c$ being the material toughness. The crack geometry $\Gamma$ is unknown, hindering minimisation of (\ref{eq:GriffithPotential}). This can be overcome by introducing a continuous phase field variable $\phi \in [0,1]$. The phase field resembles a damage variable, representing the intact state of the material when $\phi=0$ and the completely fractured state when $\phi=1$. 
A regularised potential energy can be defined considering this auxiliary phase field and a regularisation length scale $\ell$, such that (\ref{eq:GriffithPotential}) can be approximated as \citep{Bourdin2000,Borden2012}:
\begin{equation}\label{eq:RegularisedPotential}
\Pi_\ell = \int_{\Omega}\left\{ \frac{1}{2}\rho \dot{\bm{u}}\cdot \dot{\bm{u}} + \psi_s(\bm{u},c,\phi) + \psi_{chem}(\bm{u},c) + G_c \gamma(\phi) 
    \right\}\,\tm{d}V.
\end{equation}
In the regularized formulation, $\gamma$ represents a crack surface density function. In the following, we introduce a number of specific choices for the above formulation, specialising it for brittle fracture of hydrogen-containing metals. As discussed (e.g.) in Refs. \cite{Tanne2018,CST2021}, the phase field length scale can have different interpretations. As rigorously proven using $\Gamma$-convergence, the regularised functional $\Pi_\ell$ (\ref{eq:RegularisedPotential}) converges to that of $\Pi$ (\ref{eq:GriffithPotential}) for a fixed $\ell\to 0^+$ and thus minimising (\ref{eq:RegularisedPotential}) provides the solution for the Griffith variational problem. Accordingly, $\ell$ can be interpreted as a regularising parameter in its vanishing limit. However, a finite material strength is introduced for $\ell > 0^+$ and the phase field length scale becomes a material property governing the material strength, $\sigma_c$. For example, under plane stress conditions,
\begin{equation}
    \sigma_c \propto \sqrt{\frac{G_c E}{\ell}} = \frac{K_{Ic}}{\sqrt{\ell}}
\end{equation}
Thus, the fracture behaviour of the solid is governed by the material toughness $G_c$ and strength $\sigma_c$, as defined by the choice of $G_c$ and $\ell$. Therefore, phase field models provide a suitable framework to link fracture (in a Griffith sense) and damage. The phase field length scale also regularises the numerical solution; results are mesh-independent if the finite element mesh is sufficiently fine to resolve the fracture process zone, whose size is governed by $\ell$.

\subsection{Constitutive prescriptions}

\subsubsection{Linear elasticity}
\label{sec:Theory_LinEl}

We start by introducing the assumption of linear elastic behaviour. Even within the realm of brittle fracture, this is an appreciable simplification as plasticity will develop locally at the tip of sharp defects. However, the crack tip stresses predicted by conventional plasticity are insufficient to accurately predict crack tip hydrogen concentrations \cite{IJHE2016,CS2020b} and the use of strain gradient plasticity models reveals: (i) the existence of an elastic core surrounding the crack tip \cite{EJMAS2019,IJES2020}, and (ii) a crack tip stress distribution over the fracture process zone that is closer to that of linear elasticity \cite{IJSS2015,CM2017}. Thus, linear elasticity provides a conservative, less computationally demanding alternative to multi-scale plasticity models. In addition, the formulation is made under the assumptions of small strains such that the strain tensor is given by,
\begin{equation}
    \bm{\varepsilon} = \frac{1}{2}\left( \nabla\bm{u} +\nabla\bm{u}^T\right) \, .
\end{equation}

For an (undamaged) linear elastic stiffness tensor, $\mathbf{C}_0$, the strain energy density of the intact material is defined as
\begin{equation}
    \psi_0 = \frac{1}{2}\bm{\varepsilon}\cddot \mathbf{C}_0 \cddot \bm{\varepsilon}.
\end{equation}
Thus, it is assumed that the diffusive species has no influence on the strain energy, which is a common assumption for hydrogen in metals \cite{Hirth1980}.

\subsubsection{Phase field fracture}

We proceed to make constitutive choices for the phase field fracture formulation. Two models will be considered, which are known to provide accurate descriptions of fracture phenomena in a regularized setting. The strain energy density and the crack surface density function are given by:
\begin{align}
    \psi_s &= g(\phi)\psi_0 \, , \\
    \gamma &= \dfrac{1}{4c_w\ell}\left( w(\phi) + \ell^2 |\nabla\phi|^2\right) \, .
\end{align}
Where the degradation function $g(\phi)$ is continuous and monotonic and takes the values $g(0)=1$ and $g(1)=0$. The function $w(\phi)$ must fulfill $w(0)=0$ and $w(1)=1$. Finally,
\begin{equation}
    c_w = \int_0^1\sqrt{w(\varphi)} \, \tm{d}\varphi \, .
\end{equation}

This choice of phase field formulation has been shown to $\mathrm{\Gamma}$-converge to the Griffith solution \cite{Braides1998}. Common choices for $g(\phi)$ and $w(\phi)$ include $g(\phi)=(1-\phi)^2$ and $w(\phi)=\phi^2$, which produces the so-called \emph{standard} or AT2 phase field model \cite{Ambrosio1991}. Another common choice is the same quadratic degradation function and $w(\phi)=\phi$, which is often referred to as the AT1 phase field model \cite{Pham2011}. The latter formulation introduces an elastic regime prior to any damage in the solid. 

\subsubsection{Modified Fickian diffusion}
Finally, for the absorbed diffusive species, we adopt a modified version of Fickian diffusion. In addition to concentration gradients, diffusion is assumed to be driven by gradients of hydrostatic stress, such that atomic hydrogen accumulates in areas where the lattice is being expanded. Mass conservation requirements relate the rate of change of the hydrogen concentration $c$ with the hydrogen flux $\bm{J}$ through the external surface,
\begin{equation}\label{eq:Hmassbalance}
    \int_{\Omega} \frac{\tm{d} c}{\tm{d} t} \, \mathrm{d}V + \int_{\partial \Omega} \bm{J} \cdot \bm{n} \, \mathrm{d}S = 0 \, .
\end{equation}

Diffusion is driven by the gradient of the chemical potential $\nabla \mu$. Thus, for a diffusion coefficient $D$, the flux is related to $\nabla \mu$ through a linear Onsager relationship,
\begin{equation}\label{Eq:Flux}
    \bm{J}= - \frac{D c}{R T} \nabla \mu \, .
\end{equation}

The chemical potential includes a hydrostatic stress $\sigma_H$-dependent term to account for the role of volumetric strains in driving diffusion. For an occupancy of lattice sites $\theta_L$ and partial molar volume of hydrogen $\bar{V}_H$, the chemical potential of hydrogen in lattice sites is given by,
\begin{equation}\label{Eq:ChePotential}
    \mu=\mu^0 + RT \ln \frac{\theta_L}{1-\theta_L} - \bar{V}_H \sigma_H
\end{equation}

\noindent Here, $\mu^0$ denotes the chemical potential in the reference case, $R$ is the gas constant and $T$ is the absolute temperature. Consider now the relation between the occupancy and the number of sites $\theta_L=c/N$, and make the common assumptions of low occupancy ($\theta_L \ll 1$) and constant interstitial sites concentration ($\nabla N=0$); inserting (\ref{Eq:ChePotential}) into (\ref{Eq:Flux}) then renders,
\begin{equation}\label{Eq:Flux2}
   \bm{J}=-D \nabla c + \frac{Dc}{RT} \bar{V}_H \nabla \sigma_H
\end{equation}


\subsubsection{Hydrogen degradation of the toughness}

The material toughness is defined to be sensitive to the hydrogen content, $G_c (c)$, as consistently observed experimentally. First, the Langmuir–McLean isotherm is used to estimate, from the bulk concentration, the hydrogen coverage $\theta$ at decohering interfaces,
\begin{equation}\label{eq:Langmuir}
    \theta = \frac{c}{c + \exp \left( \frac{- \Delta g_b^0}{RT} \right)} \, .
\end{equation}

\noindent Here, $g_b^0$ is the binding energy for the impurity at the site of interest. A value of 30 kJ/mol is assumed throughout this study, which is representative of hydrogen trapped at grain boundaries \cite{Bhadeshia2016}. Following atomistic studies of surface energy sensitivity to hydrogen coverage \cite{Alvaro2015}, a linear degradation of the fracture energy with $\theta$ is assumed,
\begin{equation}
    G_c \left( \theta \right) = \left( 1 - \chi \theta \right) G_c (0)
\end{equation}

\noindent where $\chi$ is a hydrogen damage coefficient, a material parameter that can be estimated by calibrating with experiments \cite{CS2020} or inferred from atomistic calculations \cite{CMAME2018}. We adopt the latter approach and assume a value of $\chi=0.89$ throughout our calculations, this magnitude provides the best fit to atomistic calculations on iron \cite{CMAME2018,Jiang2004}. A failure model based on grain boundary decohesion is implicitly assumed for the above choices: a degradation law based on atomistic calculations of surface energy sensitivity to hydrogen content, and the consideration of a grain boundary binding energy in (\ref{eq:Langmuir}). However, we emphasise that the framework is universal; the degradation law can be adapted to accommodate any other mechanistic interpretation (e.g., including a dependence on parameters such as the dislocation density or the void volume fraction) or chosen to be phenomenological, relating $G_c$ to $c$ and to a hydrogen damage coefficient $\chi$ to be calibrated.

\subsection{Coupled force balances}

We proceed to present the weak and strong form of the problem considering the constitutive choices above. First, following Refs. \cite{Duda2018,Cui2021}, a scalar field $\eta$ is defined to determine the kinematics of composition changes, such that
\begin{equation}
    \dot{\eta} = \mu \, \, \, \, \, \, \text{and} \, \, \, \, \, \, \eta (\bm{x}, t) = \int_0^t \mu (\bm{x}, t) \, \text{d} t 
\end{equation}

Thus, from a kinematic viewpoint, the domain $\Omega$ can be described by the displacement $\bm{u}$, phase field parameter $\phi$, and chemical displacement $\eta$. Taking the first derivative of (\ref{eq:RegularisedPotential}) and incorporating the constitutive prescriptions adopted, the coupled weak form reads, in the absence of body forces and external tractions and fluxes, as follows:
\begin{align}\label{Eq:weak}
  \int_{\Omega} & \Bigg\{ \rho \ddot{\bm{u}} \delta \bm{u} + \left( 1 - \phi \right)^2  \bm{\sigma} : \textrm{sym} \nabla \delta \bm{u}  -2(1-\phi)\delta \phi \, \psi_0 \left( \bm{u} \right) - \frac{\textrm{d}c}{\textrm{d}t} \delta \eta \\ \nonumber
       &  + \left(\frac{Dc}{RT} \bar{V}_H \nabla \sigma_H -D \nabla c \right) \cdot \nabla \delta \eta + G_c \left( c \right) \left( \dfrac{\phi}{\ell} \delta \phi
        + \ell\nabla \phi \cdot \nabla \delta \phi \right) \Bigg\}  \, \mathrm{d}V = 0 \, .
\end{align}

The local force balances can then be readily derived by applying Gauss' divergence theorem and noting that (\ref{Eq:weak}) must hold for any kinematically admissible variations of the virtual quantities. Accordingly,
\begin{align}\label{eq:strong}
    (1 - \phi)^2 \nabla \bm{\sigma} = \, & \rho \ddot{\bm{u}} \ \nonumber \\
    G_{c} (c) \left(\frac{\phi}{\ell}  - \ell \nabla^2 \phi \right) - 2(1-\phi) \psi_{0}( \bm{u}) = \, & 0 \nonumber\\
    \frac{\textrm{d} c}{\textrm{d} t} - D \nabla^2 c +\nabla \cdot \left( \frac{D \bar{V}_H}{RT} c  \nabla \sigma_H \right) =\, & 0 
\end{align}

The coupling between the different physical elements of the problem is evident in (\ref{eq:strong}). First, as damage increases, the phase field reduces the stiffness of the solid in the linear momentum equation (\ref{eq:strong}a). As observed in (\ref{eq:strong}b), the phase field evolves driven by the competition between strain energy density $\psi_0$ and toughness $G_c$, the latter diminishing with increasing hydrogen concentration. Finally, (\ref{eq:strong}c), diffusion of atomic hydrogen is governed by concentration gradients and the lattice dilation, as characterised by the hydrostatic stress.

\section{Finite element implementation}
\label{Sec:Numerical}

We shall now describe the details of the numerical implementation in the context of the finite element method. First, some numerical considerations are presented for the phase field problem in Section \ref{Sec:Irreversibility}, to guarantee damage irreversibility and prevent crack growth from compressive stresses. Secondly, a threshold is defined in Section \ref{sec:AT1Implementation} to address the implementation peculiarities inherent to the AT1 model. Thirdly, the discretisation of the problem and the formulation of residuals and stiffness matrices is described in Section \ref{sec:FEMdis}. The implementation is carried out in the commercial finite element package Abaqus by means of a user element (UEL) subroutine. The UEL subroutine developed includes multiple choices of elements; in 2D, linear and quadratic quadrilateral elements for both plane stress and plane strain; in 3D, linear and quadratic hexahedral elements, as well as quadratic tetrahedral elements. Abaqus2Matlab is employed to pre-process the input files \cite{AES2017}.

\subsection{Addressing irreversibility and crack growth in compression}
\label{Sec:Irreversibility}

First, a history variable field $H$ is introduced to ensure damage irreversibility. Thus, for a time $t$, 
\begin{equation}
     H = \max_{\tau \in[0,t]}\psi_0( \tau). 
\end{equation}

Secondly, we introduce a strain energy decomposition to prevent cracking in compression. The volumetric-deviatoric split by Amor \textit{et al.} \cite{Amor2009} is adopted, by which the compressive volumetric strain energy does not contribute to damage. Thus, the strain energy density is decomposed into the following terms:
\begin{equation}
\psi_0^+=\frac{1}{2} K \langle tr \left( \bm{\varepsilon} \right) \rangle^2_{+} + \mu \left( \bm{\varepsilon}' : \bm{\varepsilon}' \right) \, ,
\end{equation}
\begin{equation}
\psi_0^-=\frac{1}{2} K \langle tr \left( \bm{\varepsilon} \right) \rangle^2_{-} \, ,
\end{equation}

\noindent where $K$ is the bulk modulus, $\langle a \rangle_{\pm}=(a\pm |a|)/2$ and $\bm{\varepsilon}'=\bm{\varepsilon}-tr(\bm{\varepsilon}) \bm{I} /3$. In addition, we follow the hybrid implementation of Ambati \textit{et. al} \cite{Ambati2015a} in considering only $\psi_0^+$ in the evaluation of the history variable field $H$, therefore referring to it as $H^+$ henceforth, while considering $\psi_0$ in the displacement problem.

\subsection{Implementing the AT1 phase field formulation}
\label{sec:AT1Implementation}

Unlike the AT2 phase field model, the AT1 formulation does not inherently ensure that the lower bound on the phase field is enforced. If no measures are taken, the phase field can become negative for all strains below the critical strain, which is given by
\begin{equation}
    \varepsilon_{c,AT1} = \sqrt{\dfrac{3G_c}{8\ell E}}.
\end{equation}
To overcome this, we introduce a lower bound by re-defining the history field as:
\begin{equation}
    H = \tm{max} \left[ \max_{\tau \in[0,t]}\psi_0( \tau), \, \dfrac{1}{2}E \varepsilon_{c,AT1}^2 \right] \, ,
\end{equation}

The minimum threshold employed for the history field corresponds to the strain energy density magnitude that yields a vanishing phase field in the homogeneous 1D case \cite{Wu2020c}. Other methods exist for implementing the AT1 formulation, such as using constrained optimization solvers.

\subsection{Finite element discretisation of variational principles}
\label{sec:FEMdis}

We proceed to discretise the linearised problem and present the associated residuals and stiffness matrices. First, making use of Voigt notation, the displacement field $\bm{u}$, phase field $\phi$ and hydrogen concentration $c$ can be discretised as
\begin{equation}
    \bm{u}=\sum_{i=1}^{m} \bm{N}_{i}^{\bm{u}} \bm{u}_{i} \, \, , \hspace{7mm} \phi=\sum_{i=1}^{m} N_{i} \phi_{i} \, \, , \hspace{7mm} c=\sum_{i=1}^{m} N_{i} c_{i} \, \, ,
    \centering
\end{equation}

\noindent where $N_i$ denotes the shape function associated with node $i$, for a total number of nodes $m$. Here, $\bm{N}_i^{\bm{u}}$ is a diagonal interpolation matrix with the nodal shape functions $N_i$ as components. Similarly, using the standard strain-displacement $\bm{B}$ matrices, the associated derivatives are discretised as,
\begin{equation}
    \bm{\varepsilon}= \textrm{sym} \nabla \bm{u} =\sum_{i=1}^{m} \bm{B}_{i}^{\bm{u}} \bm{u}_{i} \, \, , \hspace{7mm}  \nabla \phi=\sum_{i=1}^{m} \bm{B}_{i} \phi_{i} \, \, , \hspace{7mm} \nabla c=\sum_{i=1}^{m} \bm{B}_{i} c_{i} \, \, .
    \centering
\end{equation}

\noindent Considering this finite element discretisation and the weak form balances (\ref{Eq:weak}), the resulting discrete equations of the balances for the displacement, phase field and concentration can be expressed as the following residuals:
\begin{align}
    & \mathbf{r}_{i}^\mathbf{u}=\int_{\Omega} \left\{ \left[(1-\phi)^{2}+k\right] {(\mathbf{B}_{i}^{\mathbf{u}})}^{T} \bm{\sigma} + \rho {(\mathbf{N}_{i}^{\mathbf{u}})}^{T} \ddot{\bm{u}} \right\} \, \mathrm{d}V \\
    & r_{i}^{\phi}= \int_{\Omega} \left\{ -2(1-\phi) N_{i} \, H^+ +
    G_{c} (c) \left[\frac{\phi}{\ell} N_{i} 
    + \ell {(\mathbf{B}_{i}^{\phi})}^{T} \nabla \phi \right] \right\} \, \mathrm{d}V \\
    & r_i^c = \int_{\Omega}\left[N_i\left(\dfrac{1}{D}\frac{\textrm{d} c}{\textrm{d} t}\right) + \mathbf{B}_i^T\nabla c - \mathbf{B}_i^T\left(\frac{\overline{V}_H }{RT} c \nabla\sigma_H\right)\right] \, \text{d}V
\end{align}

\noindent where $k$ is a numerical parameter introduced to keep the system of equations well-conditioned; a value of $k=1 \times 10^{-7}$ is adopted throughout this study based on previous studies. This choice is grounded on previous studies \cite{CPB2019,CMAME2021}; the use of smaller values has no influence in the results. Subsequently, the tangent stiffness matrices are calculated as:
\begin{align}
& \bm{K}^{\mathbf{u},\mathbf{u}}_{ij} = \frac{\partial \mathbf{r}_i^\mathbf{u}}{\partial \mathbf{u}_j} = \int_\Omega \left\{ \left[\left(1-\phi\right)^2 + k \right]\left(\bm{B}^\mathbf{u}_i\right)^T\bm{C}_0\bm{B}^\mathbf{u}_j + \frac{\rho}{\left( \mathrm{d}t \right)^2} \left( \bm{N}_i^{\bm{u}} \right)^T \bm{N}_j^{\bm{u}} \right\} \,  \tm{d}V \\
& \bm{K}^{\phi,\phi}_{ij} = \frac{\partial r^\phi_i}{\partial\phi_j} = \int_\Omega \left\{ \left[ 2H^+ +\dfrac{G_c (c)}{\ell} \right] N_iN_j + G_c (c) \ell \bm{B}_i^T\bm{B}_j\right\}\, \tm{d}V\\
& \bm{K}^{c,c}_{ij} =\frac{\partial r^c_i}{\partial c_j} =\int_\Omega\left(N_i^T \dfrac{1}{D \mathrm{d}t}N_j + \bm{B}_i^T\bm{B}_j - \bm{B}_i^T\dfrac{\overline{V}_H}{RT}\nabla\sigma_H N_j\right)\,\tm{d}V
\end{align}

Unless otherwise stated, the global system of equations for the linearised problem will be solved using a staggered, alternative minimisation scheme \cite{Miehe2010,CMAME2018}. 

\section{Results}
\label{Sec:Results}

The potential of the formulation in simulating complex fracture phenomena and transforming engineering assessment is demonstrated by addressing four case studies of particular interest. First, we model for the first time dynamic failure of a hydrogen pre-charged steel plate, using the AT1 model; see Section \ref{sec:DynamicBranching}. Next, in Section \ref{Sec:TiltedCrack}, we address the failure of a tensile bar due to the interaction between a tilted crack and a neighboring void. Thirdly, in Section \ref{sec:CaseStudy3}, we model the brittle fracture of an anchor in a concrete element subjected to a corrosive environment, following the ASTM E488 standard \cite{ASTME488}. Finally, the progressive failure of a pipeline is simulated; coupling modelling with in-line inspection data, the model incorporates the numerous defects that typically arise due to pitting corrosion and captures their growth and coalescence under in-service conditions.

\subsection{Crack branching in an embrittled steel plate due to dynamic loading}
\label{sec:DynamicBranching}

\begin{figure}[H]
    \centering
    \includegraphics{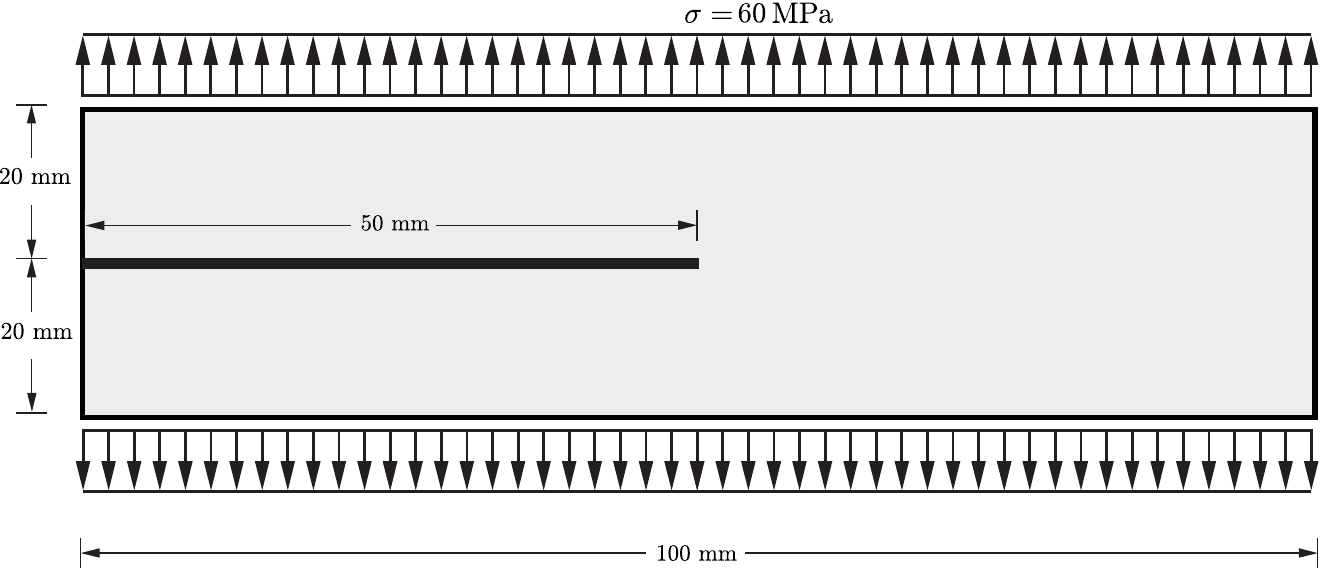}
    \caption{Dimensions and loading conditions for the dynamic branching problem.}
    \label{fig:BranchSketch}
\end{figure}

For the first case study, we consider the paradigmatic boundary value problem of dynamic crack branching in a rectangular plate \cite{Song2008,Borden2012,Zhou2018a}. The geometry and boundary conditions are given in Fig. \ref{fig:BranchSketch}. This well-known case study is based on dynamic experiments on brittle materials such as glassy polymers \cite{Ramulu1985}. Here, we aim to illustrate the influence of hydrogen on the dynamic fracture pattern of a martensitic steel of type 440C, which is known to exhibit very brittle fracture in the presence of hydrogen \cite{Jewett1973}.

The elastic parameters for the martensitic steel considered are $E=194$ GPa, $\nu = 0.3$ and $\rho=7850$ kg/m$^3$, yielding a Rayleigh wave speed of $v_n = 5768$ m/s. The plane strain fracture toughness of martensitic steel of type 440C is $K_{IC} \approx 22$ MPa$\cdot\tm{m}^{1/2}$ \citep{Lou1983}, which corresponds to a Griffith energy of $G_c=2.33$ J/$\tm{mm}^2$; recall that $K_{Ic}=\sqrt{G_c E/(1-\nu^2)}$. The material is assumed to be pre-charged with either 5 or 0.1 wppm of hydrogen. As a reference case, a specimen without hydrogen has also been included. The fracture length scale has been chosen as two times the characteristic element length $\ell = 2h_e = 0.4$ mm and the specimen is subjected to a tensile impact load of $60$ MPa. The domain is meshed with quadratic rectangular (square) elements and the initial crack has been introduced through the phase field rather than as a discontinuity in the mesh. A total of 50,000 elements are used, taking advantage of symmetry. In this dynamic case, crack propagation is expected to be faster than hydrogen diffusion by at least an order of magnitude, allowing us to neglect the influence of hydrogen transport. Without the diffusion equations the global stiffness matrix is symmetric, which enables us to make use of a monolithic scheme in conjunction with the quasi-Newton method \cite{Wu2020a,TAFM2020} to solve the problem more efficiently. For this specific problem, the AT1 phase field model is used. Thus, this case study also constitutes the first example involving AT1 and quasi-Newton solution methods, confirming the successful performance observed with other phase field models \cite{Wu2020a,TAFM2020}. The increment size has been chosen as $\Delta t < h_e/v_n$ to completely resolve the stress waves in the material.

\begin{figure}[H]
    \centering
    \includegraphics[width=0.8\textwidth]{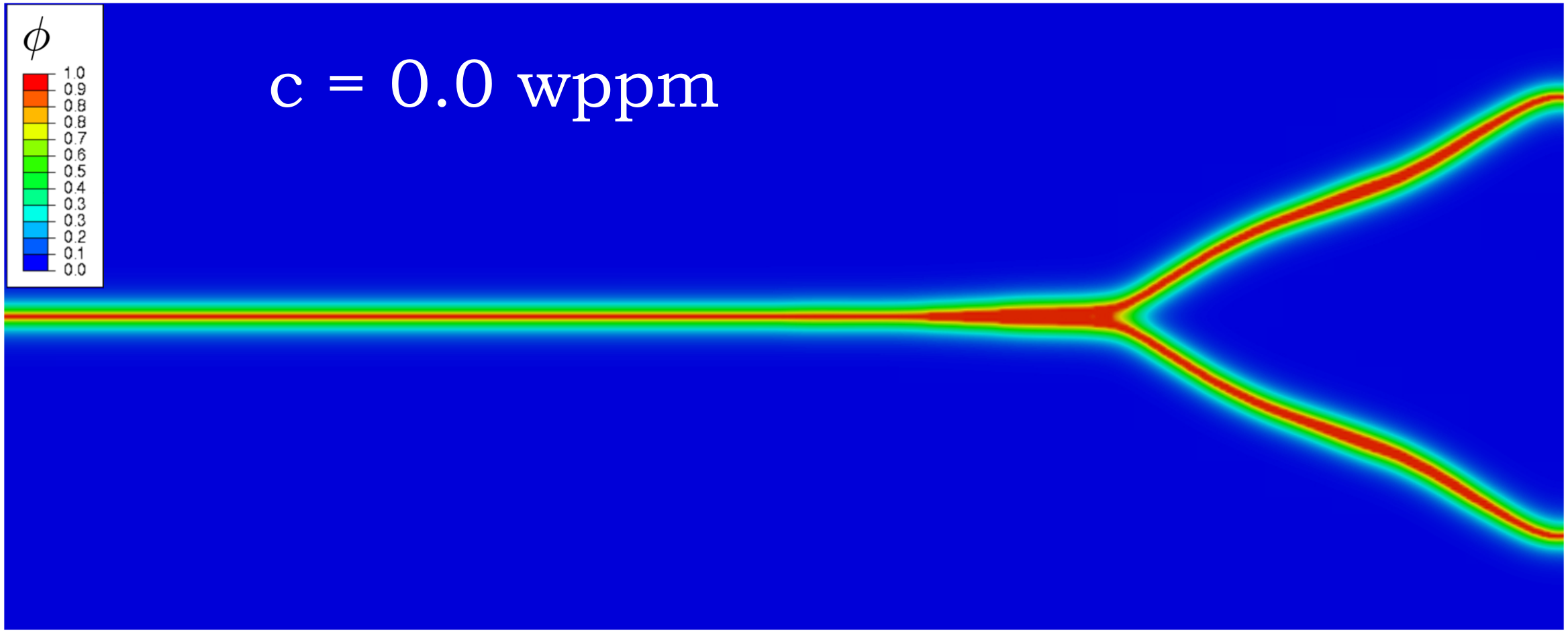}
    \includegraphics[width=0.8\textwidth]{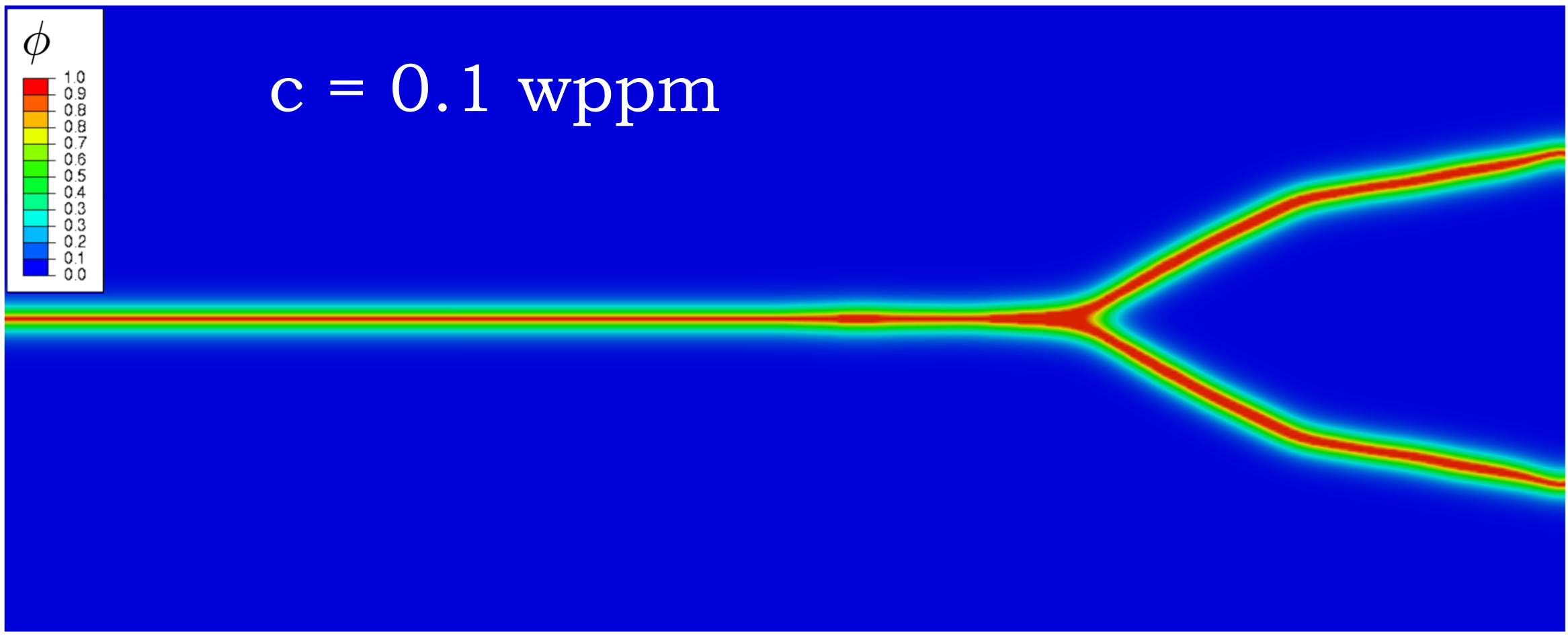}     \includegraphics[width=0.8\textwidth]{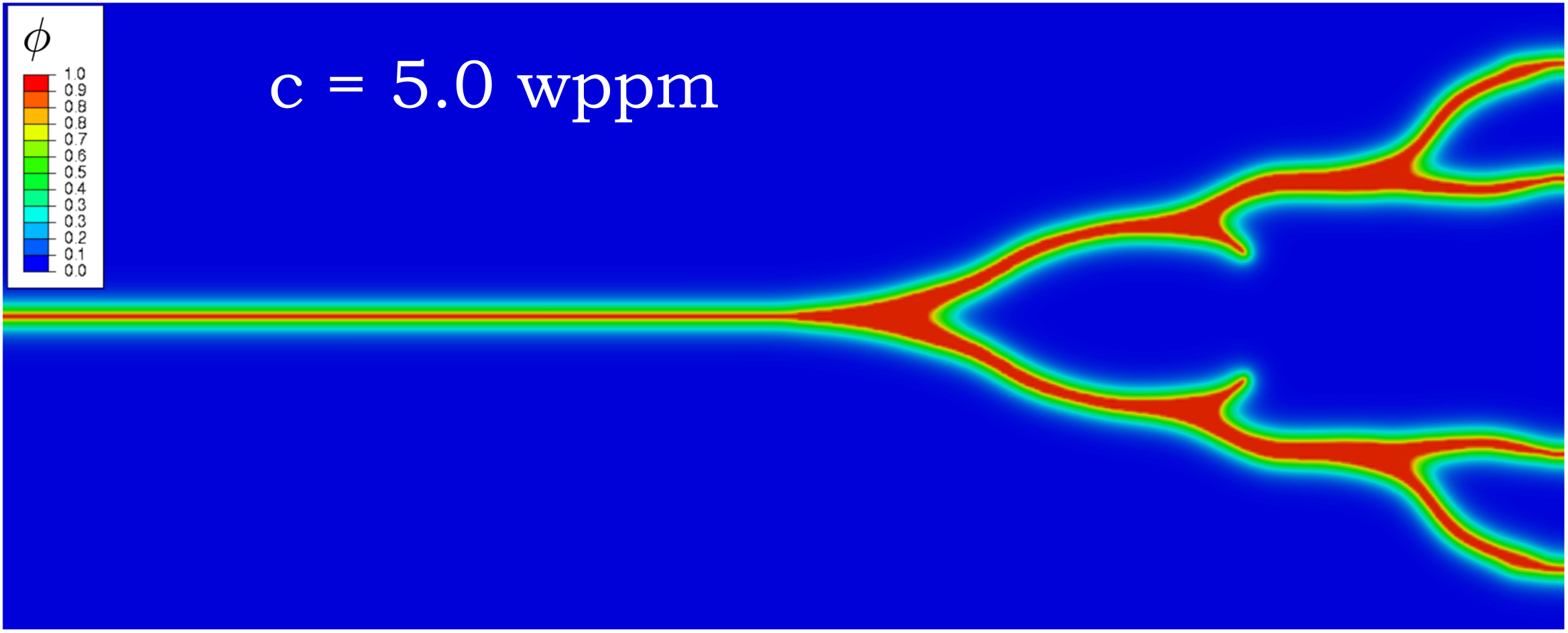}
    \caption{Increasing degree of crack branching with increasing hydrogen pre-charged content in a martensitic steel plate subjected to dynamic loading. Phase field contours, with blue representing intact material ($\phi \approx 0$) and red denoting the crack ($\phi \approx 1$).}
    \label{fig:Branching}
\end{figure}

The finite element results obtained are reported in Fig. \ref{fig:Branching} in terms of the phase field contours; blue colour represents intact material ($\phi \approx 0$) while red colour is used to denote the crack ($\phi \approx 1$). The results reveal that the crack pattern is very sensitive to the hydrogen content. A noticeable influence is observed for the case of a $c_0=0.1$ wppm hydrogen concentration and extensive branching is seen for the case of $c_0=5$ wppm, which might be interpreted as mild shattering. We emphasise that branching is intrinsically related to inertia effects, which are significant in this boundary value problem, and the material brittleness, which is sensitive to the hydrogen content. The results showcase the capability of the proposed framework for capturing inertia effects and the complex crack patterns that might occur in hydrogen-assisted fractures. The remaining case studies will deal with quasi-static loading conditions and accordingly the kinetic energy terms will be dropped from the model. 

\subsection{Void-crack interaction in a 3D tensile bar}
\label{Sec:TiltedCrack}

\begin{figure}[H]
\centering
\includegraphics[width=0.7\textwidth]{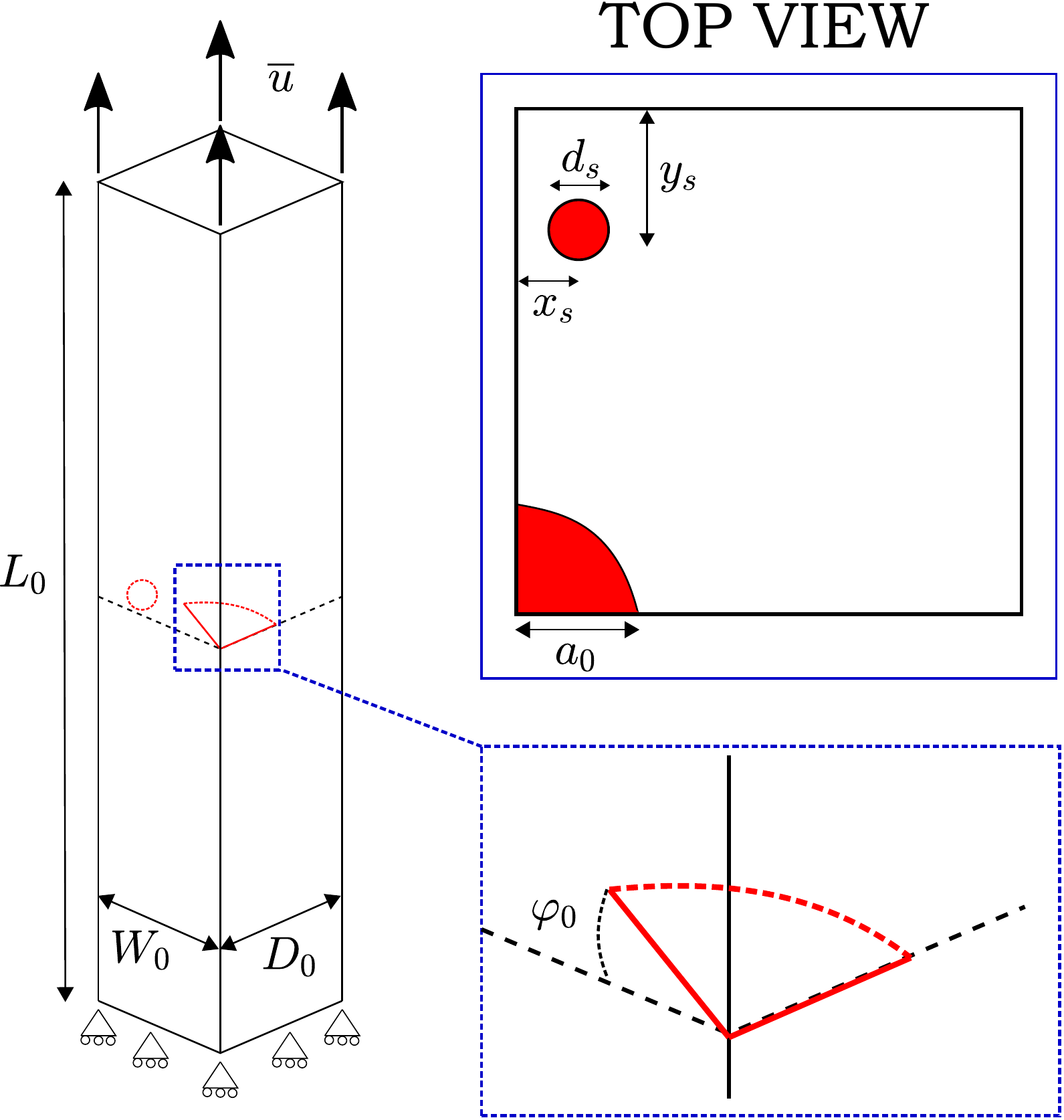}
\caption{Void-crack interaction in a 3D tensile bar: geometry and loading configuration.}
\label{fig:TinySpecSketch}
\end{figure}

The second case study deals with the prediction of crack initiation, growth and unstable failure in a rectangular prismatic bar containing a circular void and a tilted circular edge crack. The bar is shown in Fig. \ref{fig:TinySpecSketch}, with length $L_0 = 800$ mm and cross-sectional dimensions $W_0 = D_0 = 80$ mm. The bar contains an edge crack of radius  $a_0=15$ mm, which is tilted $\varphi_0=20^{\circ}$. In addition, a spherical void of diameter $d_s = 8$ mm exists in the plane of the crack. The void has been pre-charged with a hydrogen content of 1 wppm. Neumann-type boundary conditions $\bm{J}=\bm{0}$ are considered in all the outer boundaries. The in-plane position of the void is illustrated in Fig. \ref{fig:TinySpecSketch}, with $x_s=10$ mm and  $y_s=20$ mm. The bar is simply supported at one end and subjected to tension at the other end by means of a prescribed displacement $\overline{u}$.

The void and crack are introduced as initial conditions in the phase field. The domain is meshed with quadratic tetrahedrons with a characteristic element size of $h_e=1$ mm. We consider a high strength steel with $E=210$ GPa, $\nu=0.3$, $G_c = 4.54$ kJ/m$^2$ and $\ell=8$ mm. The phase field AT2 model is considered and the material is assumed to have a diffusion coefficient of $D=0.0127$ mm$^2$/s. A small loading rate is applied, allowing for the hydrogen to re-distribute through the sample.

\begin{figure}[H]
    \centering
    \begin{minipage}{0.9\linewidth}
    \begin{center}
          \includegraphics[width=\linewidth]{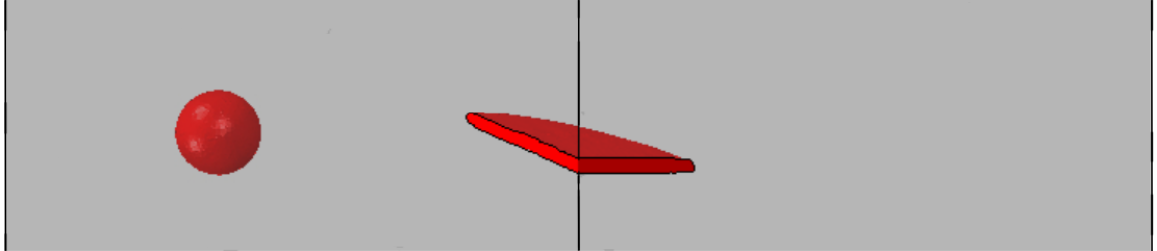}\vspace{-1.6cm}
    \end{center}
   \hspace{0.5cm}(a)
\end{minipage}\\[0.4cm]
    \begin{minipage}{0.9\linewidth}
    \begin{center}
       \includegraphics[width=\linewidth]{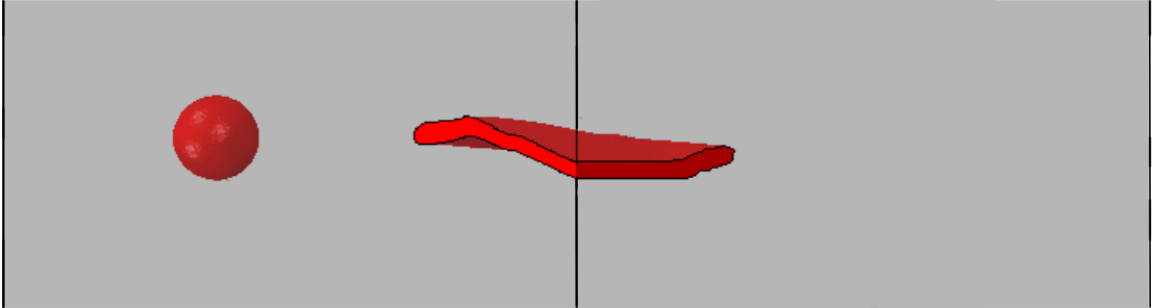}\vspace{-1.6cm}
    \end{center}
 \hspace{0.5cm}(b) 
\end{minipage} \\[0.4cm]
    \begin{minipage}{0.9\linewidth}
    \begin{center}
        \includegraphics[width=\linewidth]{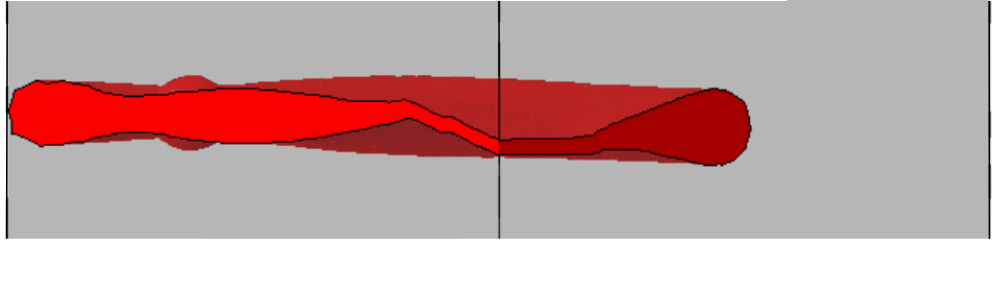}\vspace{-2cm}
    \end{center}
 \hspace{0.5cm} (c)
    \end{minipage}
    \caption{Crack initiation and growth in a rectangular prismatic bar with multiple defects, side view. The bar is subjected to tension using a prescribed displacement of: (a) $\overline{u}=0$ mm, (b) $\overline{u}=6.53$ mm, (c) $\overline{u}=6.57$ mm.}
    \label{fig:TinySpecSide}
\end{figure}

\begin{figure}[H]
    \centering
    \begin{minipage}{0.45\linewidth}
    \centering
    \includegraphics[width=\linewidth]{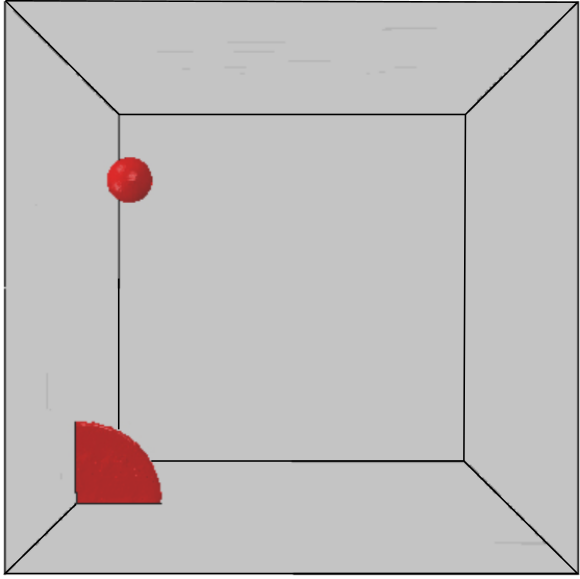}\vspace{-1cm}
    (a)
\end{minipage}
    \begin{minipage}{0.45\linewidth}
    \centering
    \includegraphics[width=\linewidth]{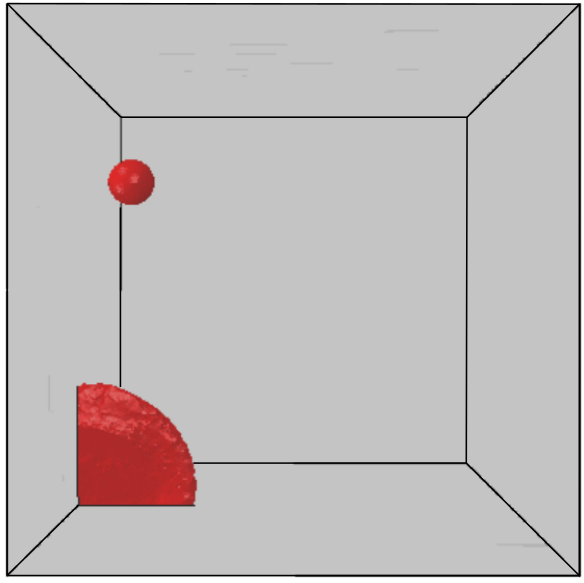}\vspace{-1cm}
    (b)
\end{minipage} \\[0.2cm]
    \begin{minipage}{0.45\linewidth}
    \centering
    \includegraphics[width=\linewidth]{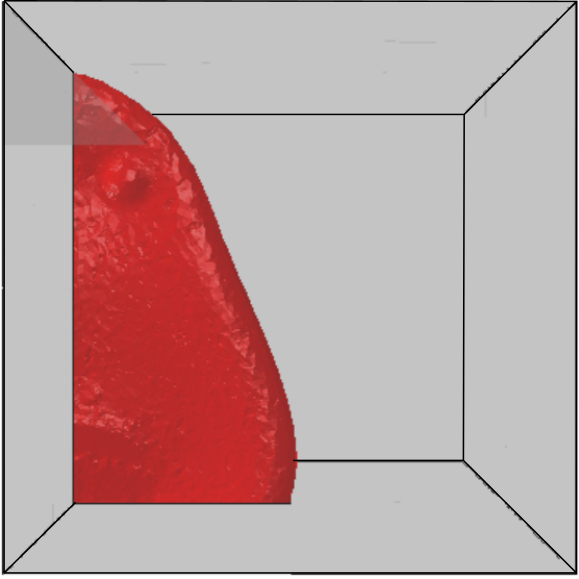}\vspace{-1cm}
    (c)
\end{minipage}
    \begin{minipage}{0.45\linewidth}
    \centering
    \includegraphics[width=\linewidth]{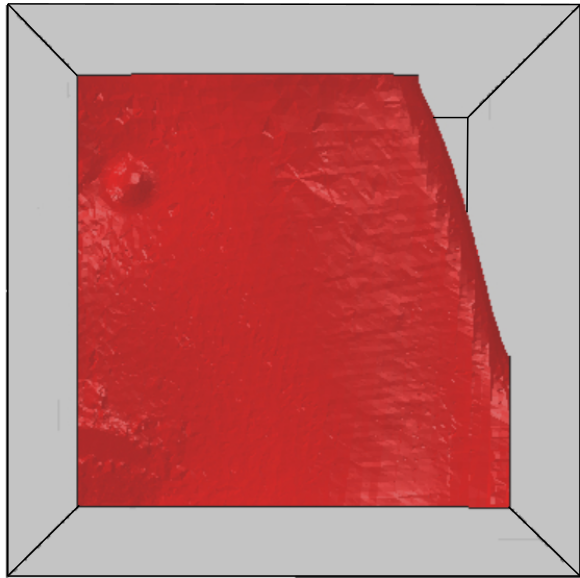}\vspace{-1cm}
    (d)
\end{minipage}
    \caption{Crack initiation and growth in a rectangular prismatic bar with multiple defects, top view. (a) Initial configuration, (b) initial crack kinking, and (c)-(d) stages of unstable crack propagation.}
    \label{fig:TinySpecTop}
\end{figure}

The results obtained are shown in Figs. \ref{fig:TinySpecSide} (side view) and \ref{fig:TinySpecTop} (top view). Cracking contours ($\phi>0.95$) are shown for different levels of remote loading. Initially, the crack kinks towards a plane perpendicular to the applied load, but it preserves its bluntness - see Figs. \ref{fig:TinySpecSide}(a)-(b). After the reorientation of the tilted crack, the interaction with the stress concentration of the void is sufficient for the crack to propagate in an unstable manner, as shown in Figs. \ref{fig:TinySpecSide}(c) and \ref{fig:TinySpecTop}(c)-(d). The stages of unstable crack growth, occurring almost instantaneously, can be captured with a staggered scheme and small time increments. As shown in Figs. \ref{fig:TinySpecTop}(c)-(d), the crack first interacts with the void and propagates up to the failure of the entire cross-section. The hydrogen in the void diffuses to embrittle, mainly, three regions: (i) the entire crack front, (ii) the region surrounding the void, due to the associated stress concentration, and (iii) the diffusion path between the void and crack tip. The accumulation of hydrogen in these regions facilitates the crack-void interaction, resulting in the cracking pattern observed. \\

\subsection{Virtual experiments: design of screw anchors against brittle fracture}
\label{sec:CaseStudy3}

Our third case study models the failure of screw anchors in aggressive environments, mimicking the testing conditions of the ASTM E488/E488M standard \cite{ASTME488}. The aim is to showcase the capabilities of the model in optimising experimental campaigns and certification, while addressing the phenomenon of bolt failure due to hydrogen ingress, a significant concern in offshore engineering \cite{Wolfe1990}.\\

\begin{figure}[H]
\centering
\includegraphics[width=1.2\textwidth]{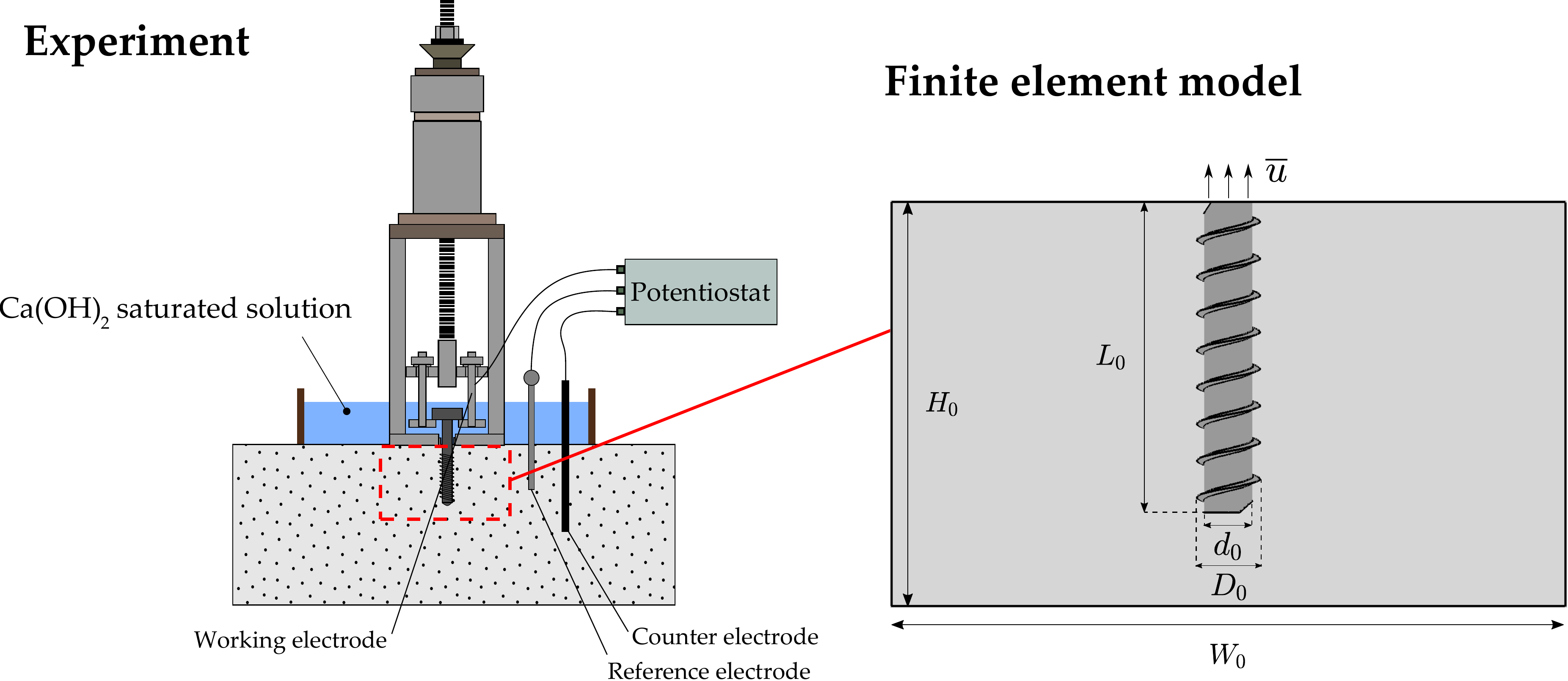}
\caption{Virtual Testing of screw anchors. Confined test setup and sketch of the finite element model.}
\label{fig:ScrewSketch}
\end{figure}

Screw anchors for use in concrete are often made out of high-strength galvanized steel. The zinc coating provides excellent corrosion protection but it can potentially increase the risk of hydrogen embrittlement. If the coating is damaged, the corrosion potential can be lowered sufficiently for hydrogen evolution to occur even in the highly alkaline conditions observed in concrete. To simulate the brittle failure of screws exposed to hydrogen-containing environments, we chose to replicate the test provided in the ASTM E488/E488M standard \cite{ASTME488} for assessing the susceptibility of screw anchors to hydrogen embrittlement. The setup for the test is illustrated in Fig. \ref{fig:ScrewSketch}. The general approach for the standardised test is to pre-charge the anchor with hydrogen and then carry out a tensile test up to ultimate failure. The pre-charging occurs by exposing the sample, for an extended period of time, to a solution representative of the one found in concrete pores. The pre-charging is carried out with the anchor in tension and potentiostatic control of the potential. The applied potential is kept sufficiently low for hydrogen evolution to occur. To estimate the hydrogen content of the bolt we consider the work by Recio \textit{et al.} \cite{Recio2011}, where a similar pre-charging protocol was employed. Based on the results therein and the differences between protocols, the initial hydrogen concentration is expected to be equal to 2 wppm or higher. A magnitude of 2 wppm is assumed.

\begin{figure}[H]
    \centering
    \begin{minipage}{0.37\textwidth}
    a)
    \includegraphics[width=\textwidth]{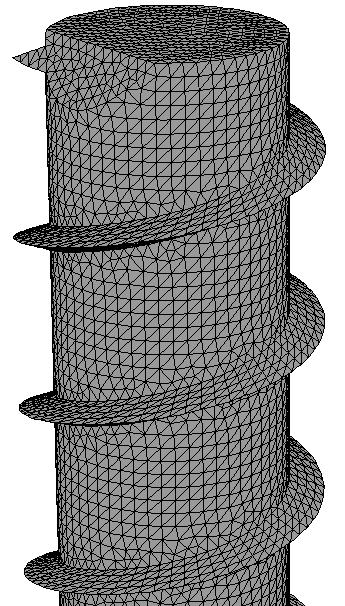}
    \end{minipage}
        \begin{minipage}{0.57\textwidth}
        b)
    \includegraphics[width=\textwidth]{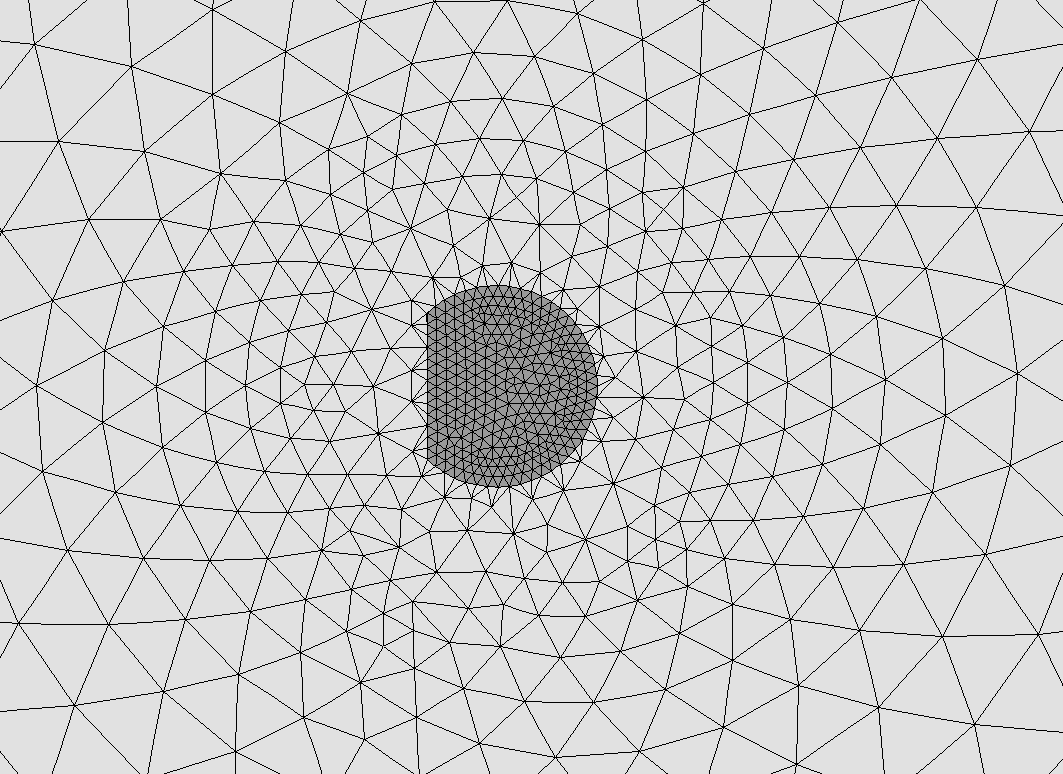}
    \end{minipage}
    \caption{Finite element mesh for the screw anchor test setup: a) mesh of the screw anchor; and, b) mesh of the concrete (light grey) around the screw anchor (dark grey), as seen from above.}
    \label{fig:ScrewMesh}
\end{figure}

The modelling domain consists of a rectangular concrete slab and a steel screw anchor, as illustrated in Fig. \ref{fig:ScrewSketch}. The concrete slab has height $H_0=76.2$ mm and the width in both directions is $W_0= 127$ mm. The screw anchor has a length $L_0= 58.4$ mm and a core diameter of $d_0=9.1$ mm. The outer diameter of the thread is $D_0=11.9$ mm. Only mechanical deformation is considered in the concrete slab, which is modelled as a linear elastic material, while the steel anchor uses the deformation-diffusion-damage formulation presented above. Contact between the thread surface of the screw anchor and the concrete is modelled including friction, with a coefficient of friction $\mu=0.35$. Material constants for the concrete slab are $E_c = 23.6$ GPa and $\nu_c=0.2$. For the steel screw, the material properties read: $E_s=210$ GPa, $\nu_s=0.3$ and $G_c = 64$ N/mm$^2$, corresponding to a plane strain fracture toughness of $121.5$ MPa$\cdot \tm{m}^{1/2}$. The diffusion coefficient is assumed to be $D_s = 0.0127$ mm$^2$/s. The phase field length scale has been set to $\ell = 3.05$ mm, approximately five times the characteristic element length $h_e=0.6$ mm. Both parts of the modeling domain are meshed using tetrahedral elements with quadratic shape functions. The mesh is illustrated in Fig. \ref{fig:ScrewMesh}. The concrete domain is discretised with 117,456 while a total of 155,278 elements are used for the screw. As the problem considers crack initiation from a stress concentration, it is deemed suitable to adopt the AT1 phase field model \cite{Tanne2018}. The standard test method contains no specifications about the load rate, although it can be expected to have an influence on the result. In the model presented here, the screw anchor fractures after a loading period of $0.53$ s. The load is applied under displacement control conditions.

\begin{figure}[H]
\centering
\includegraphics[width=0.7\textwidth]{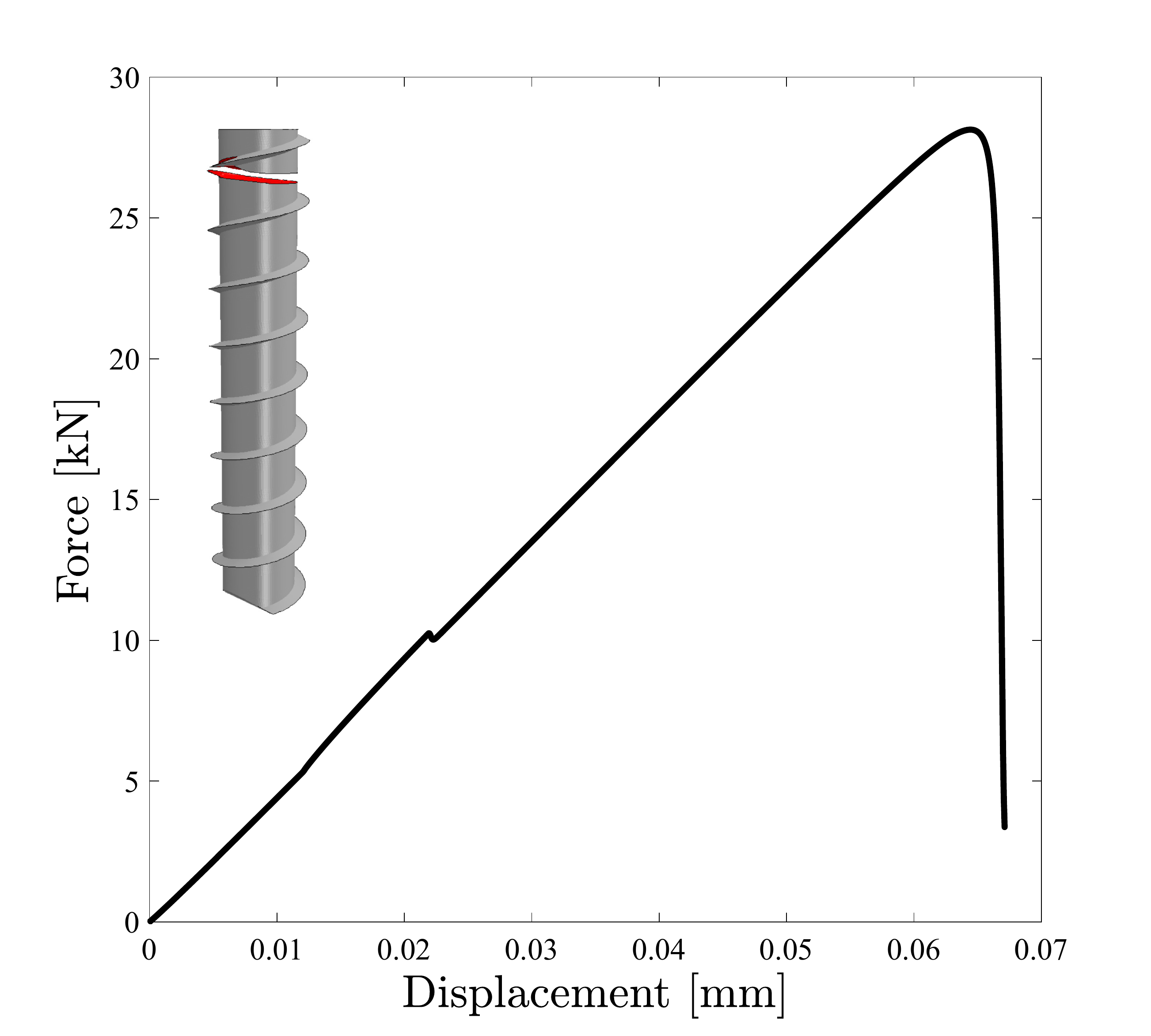}
\caption{Virtual Testing of screw anchors. Force versus displacement curve for the screw test.}
\label{fig:ScrewFD}
\end{figure}

\begin{figure}[H]
    \centering
    \begin{minipage}{0.47\textwidth}
    a)
    \includegraphics[width=\textwidth]{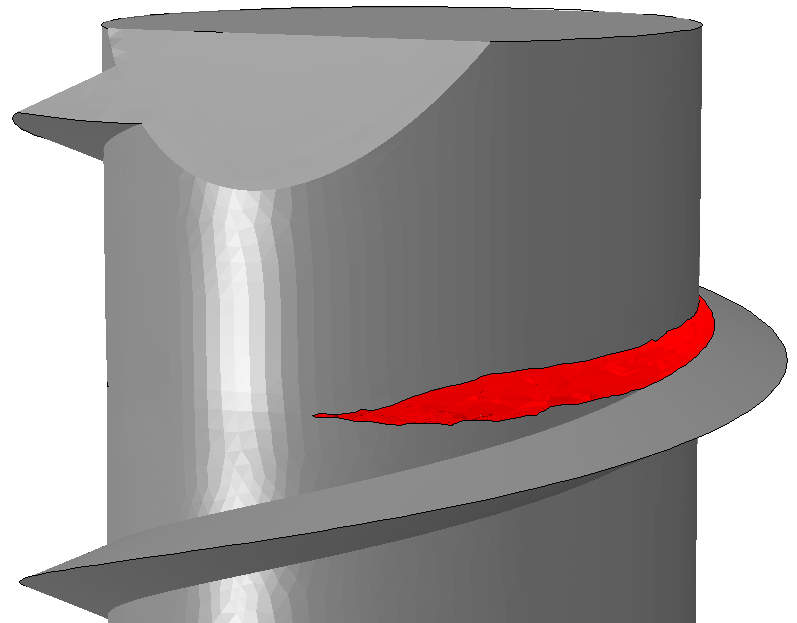}
    \end{minipage}
        \begin{minipage}{0.47\textwidth}
        b)
    \includegraphics[width=\textwidth]{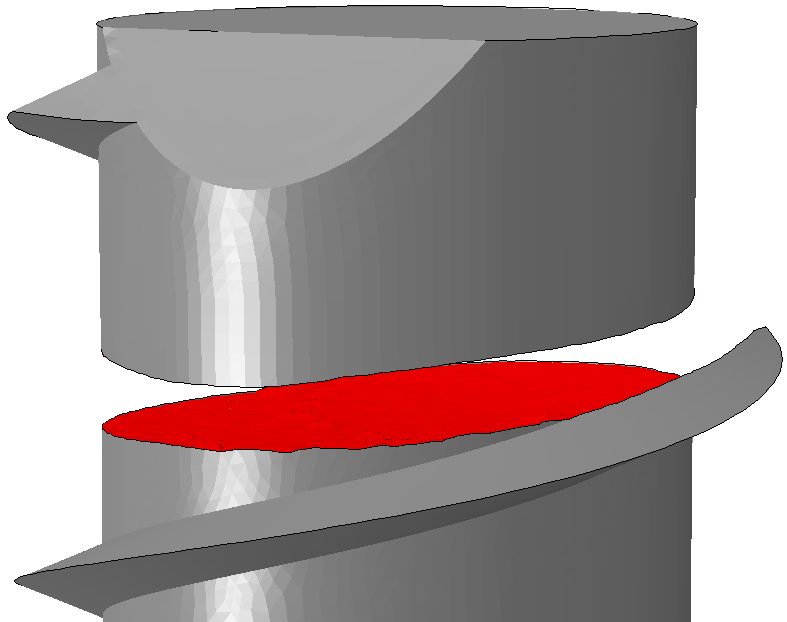}
    \end{minipage}
    \caption{Virtual Testing of screw anchors. Broken state of the screw anchor: a) areas with $\phi > 0.98$ removed, $\phi=0.98$ shown in red, and b)  $\phi > 0.96$ removed, $\phi=0.96$ shown in red.}
    \label{fig:ScrewResult}
\end{figure}

Representative finite element results are shown in Figs. \ref{fig:ScrewFD} and \ref{fig:ScrewResult}. The force versus displacement response is shown in Fig. \ref{fig:ScrewFD}; the force increases until it reaches a peak value of $27.88$ kN, after which a significant drop is observed, indicative of unstable brittle fracture. The peak force corresponds to a nominal core stress of $429$ MPa, which is less than the typical yield stress of the materials used in these applications. The broken state of the screw is shown in Fig. \ref{fig:ScrewResult}, where it can be seen that fracture occurs close to the head of the screw. A stress concentration is observed along the root of the thread, where crack initiation is observed - see Fig. \ref{fig:ScrewResult}a. The location of crack initiation agrees with expectations as the first winding of the thread carries the highest load. The predictions of the proposed model indicate that a galvanized screw anchor of this type is susceptible to brittle fracture under hydrogen embrittlement if the steel is exposed to a concrete pore solution.

\subsection{Coupling with NDE: Failure of a pipeline with internal defects}
\label{sec:Pipeline}

\begin{figure}[H]
    \centering
    \includegraphics[width=\linewidth]{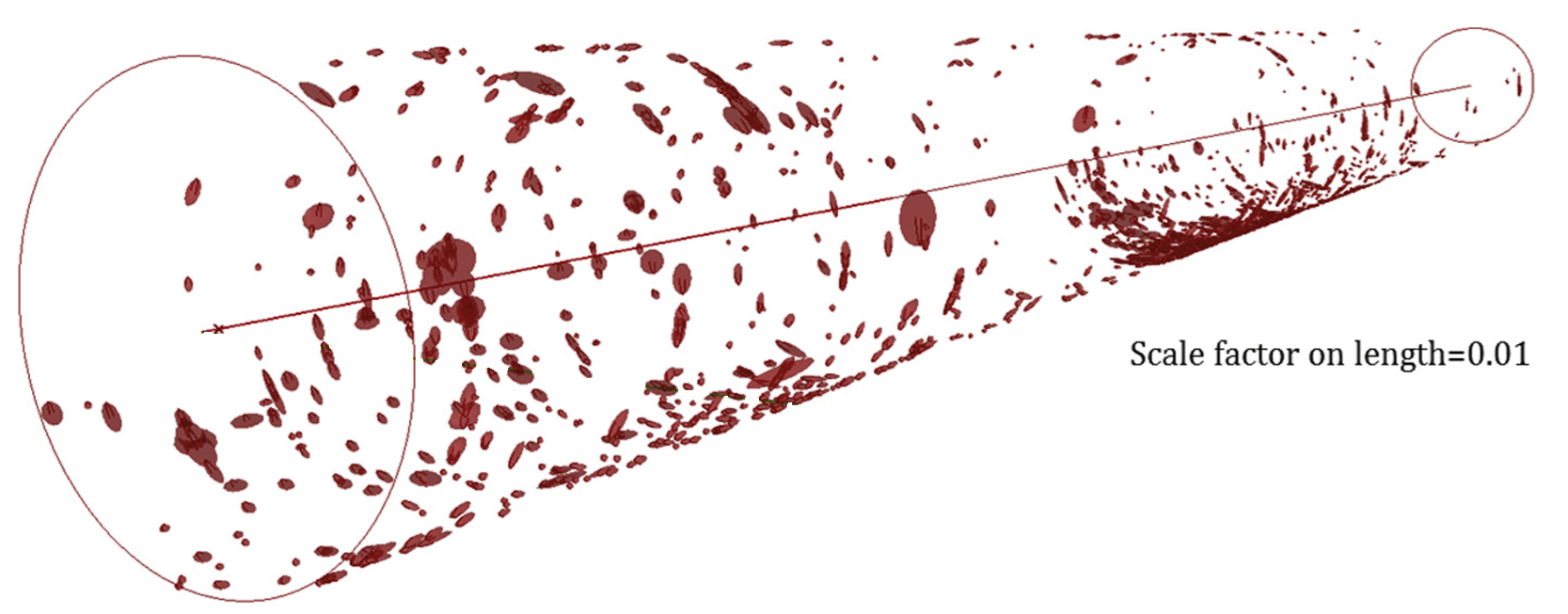}
    \caption{Initial distribution of defects in a corroded pipeline, as measured by in-line inspection (ILI). Adapted from \citep{Larrosa2018}.}
    \label{fig:PipeInitial}
\end{figure}

The final case study showcases how phase field methods can be coupled to non-destructive evaluation (NDE) to develop high-fidelity models for real-time assessment of critical infrastructure. This is exemplified with the modelling of the progressive failure of a pipeline with numerous internal defects. The structural integrity of pipelines subjected to aggressive environments is a major concern in the energy industry \cite{Heidersbach2018}. Corrosion damage nucleates pits that act as stress concentrators, attracting hydrogen that triggers early cracking. Thus, we consider the structural failure of a pipeline as a paradigmatic case study and show how our multi-physics model can predict cracking in large scale components as a function of the material properties, the environmental hydrogen concentration and the loading conditions. In addition, we aim at enabling a \emph{Virtual Testing} paradigm by coupling the present phase field predictions with defect characterisation from NDE inspections. The distribution of defects in a pipeline is shown in Fig. \ref{fig:PipeInitial}, as measured from in-line inspection (ILI) and reported by Larrosa \textit{et al.} \cite{Larrosa2018}. A three-dimensional finite element model is developed based on this data.

\begin{figure}[H]
    \centering
    \begin{minipage}{0.45\linewidth}
    \centering
    \includegraphics[width=\linewidth]{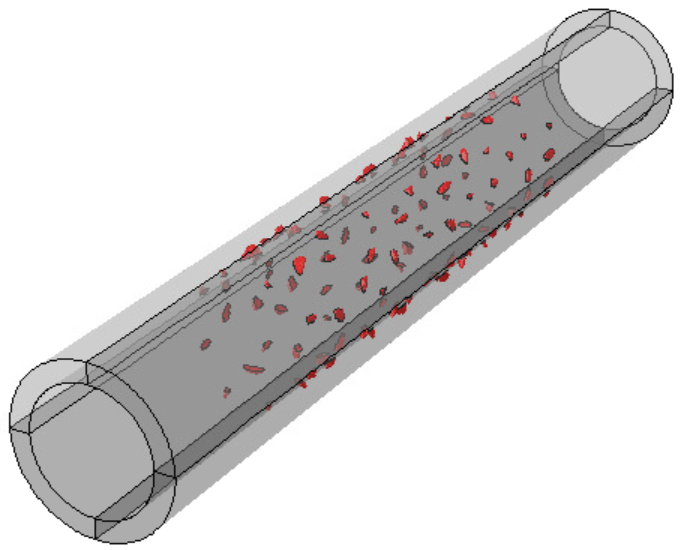}\vspace{-1cm}
    (a)
\end{minipage}
    \begin{minipage}{0.45\linewidth}
    \centering
    \includegraphics[width=\linewidth]{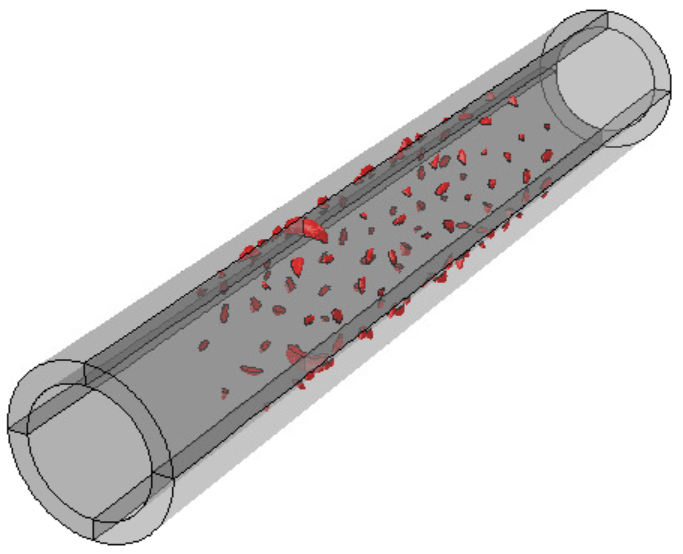}\vspace{-1cm}
    (b)
 \end{minipage}   \\
    \begin{minipage}{0.45\linewidth}
    \centering
    \includegraphics[width=\linewidth]{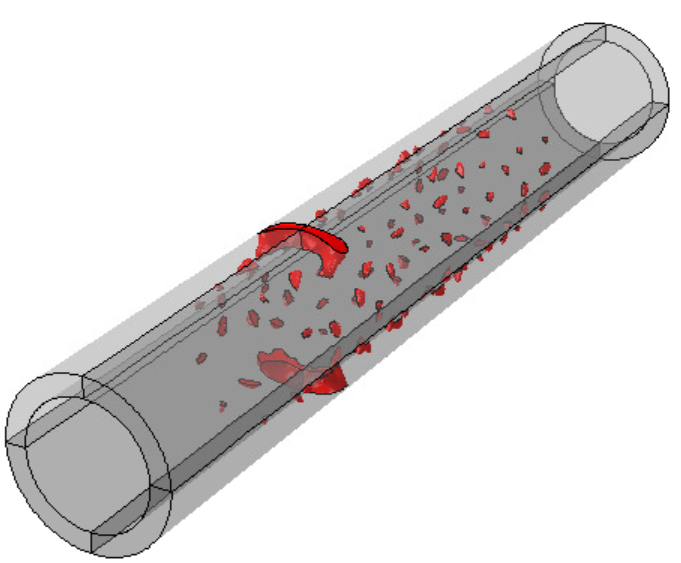} \vspace{-1cm}
    (c)
\end{minipage} 
    \begin{minipage}{0.45\linewidth}
    \centering
    \includegraphics[width=\linewidth]{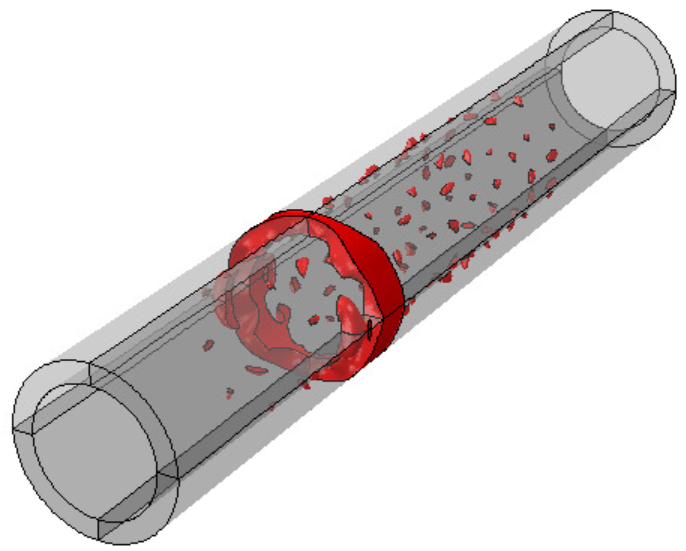}\vspace{-1cm}
    (d)
\end{minipage} \\
    \begin{minipage}{0.45\linewidth}
    \centering
    \includegraphics[width=\linewidth]{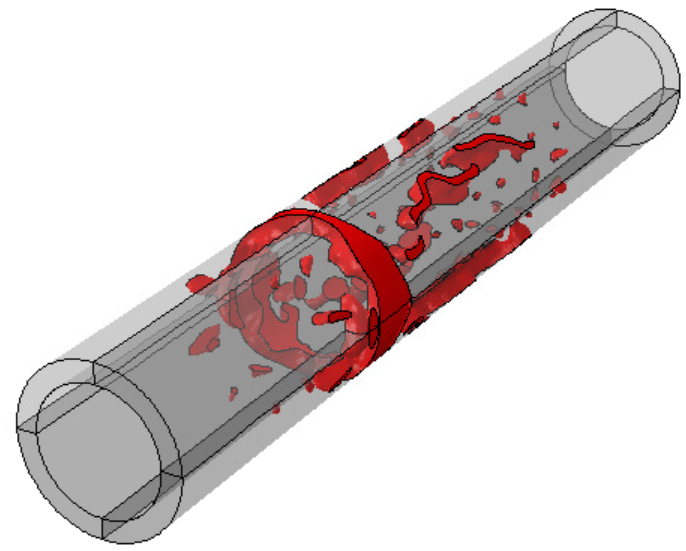}\vspace{-1cm}
    (e)
\end{minipage}
    \begin{minipage}{0.45\linewidth}
    \centering
    \includegraphics[width=\linewidth]{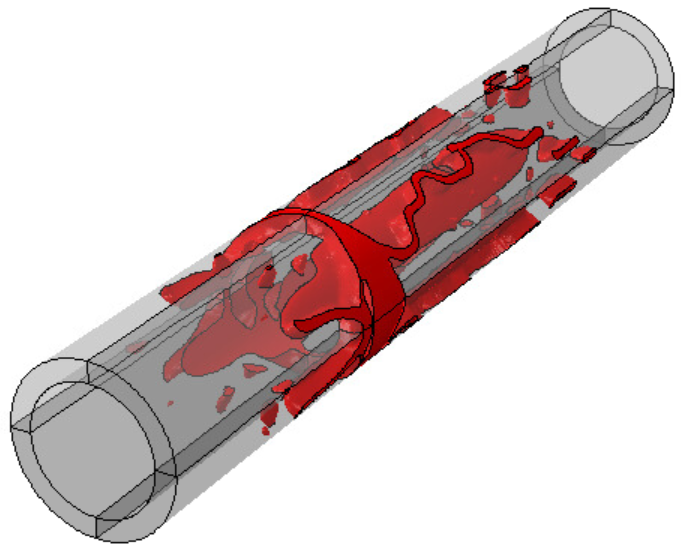}\vspace{-1cm}
    (f)
\end{minipage}
    \caption{Crack initiation and growth predicted in a pipeline with an initial distribution of defects. The figure shows a transparent cross-section of the pipeline with $\phi=0.8$ contours. The pipe is subjected to constant tension of 105 MPa, although this load is no longer carried when the crack severs the pipe in (d). The internal pressure in the pipe is (a) 0 MPa, (b) 19.6 MPa, (c) 22.1 MPa, (d) 23.8 MPa, (e) 75.7 MPa, and (f) 85 MPa.}
    \label{fig:Pipeline}
\end{figure}

A total of 1,750 defects were characterised by Larrosa \textit{et al.} \cite{Larrosa2018} in the pipeline shown in Fig. \ref{fig:PipeInitial}, of length 11 km, outer radius 162 mm, and thickness 40 mm. We mimic the outer radius and thickness of the pipeline but we restrict our attention to a critical span of 2 meters length, containing a total of 112 defects. Taking advantage of symmetry, only a quarter of the pipe cross-section is modelled. A uniform mesh of 129,600 20-node brick elements is employed, with the characteristic element length being equal to $h=4$ mm. The internal pit defects are introduced by prescribing $\phi=1$ as an initial condition. The defects are approximated as ellipsoids and a script is created to identify nodal sets based on the defect location, dimensions and rotations. The depths, lengths and widths of the defects are taken from a normal distribution that follows the mean and standard deviation reported from the in-line inspection (ILI) rendering. The defects are also subjected to a random rotation. Material properties are given by $E=210$ GPa, $\nu=0.3$, $\ell=8$mm, $D=0.0127$ mm$^2$/s, and $G_{c}=140$ kJ/m$^2$ (as estimated from a fracture toughness of $K_{Ic}=180$ MPa$\sqrt{m}$). The phase field AT2 model is considered. We assume that the pipeline has been pre-charged with an initial hydrogen concentration of $c_0=1$ wppm and is continuously exposed to the same environment, which corresponds to a 3\% NaCl aqueous solution. We note that a fixed concentration is prescribed for simplicity but a constant chemical potential should instead be used as boundary condition when aiming at quantitative results \cite{DiLeo2013,IJHE2016,Diaz2016b}. Mimicking in-service loading conditions in a riser, we subject the pipeline to an internal pressure of 152 MPa and to axial tension, with a remote stress of 105 MPa. The internal pressure is increased linearly in time while the remote stress is held constant throughout the analysis. The evolution of defects and cracks predicted is shown in Fig. \ref{fig:Pipeline} by plotting the completely cracked regions ($\phi>0.8$).

\begin{figure}[H]
    \centering
    \begin{minipage}{0.8\linewidth}
    \begin{center}
          \includegraphics[width=\linewidth]{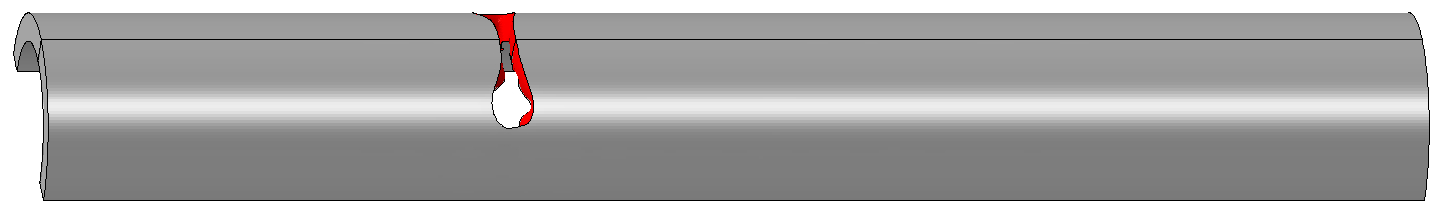}
    \end{center}
   \hspace{0.5cm}(a)
\end{minipage}\\[0.2cm]
    \begin{minipage}{0.8\linewidth}
    \begin{center}
       \includegraphics[width=\linewidth]{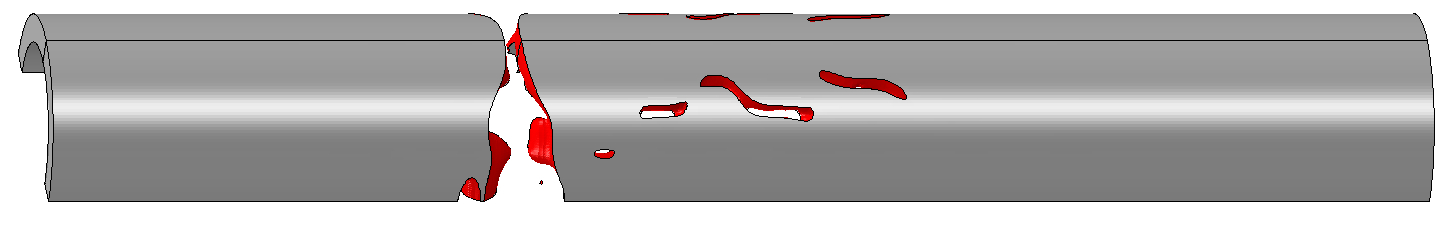}
    \end{center}
 \hspace{0.5cm}(b) 
\end{minipage} \\[0.2cm]
    \begin{minipage}{0.8\linewidth}
    \begin{center}
        \includegraphics[width=\linewidth]{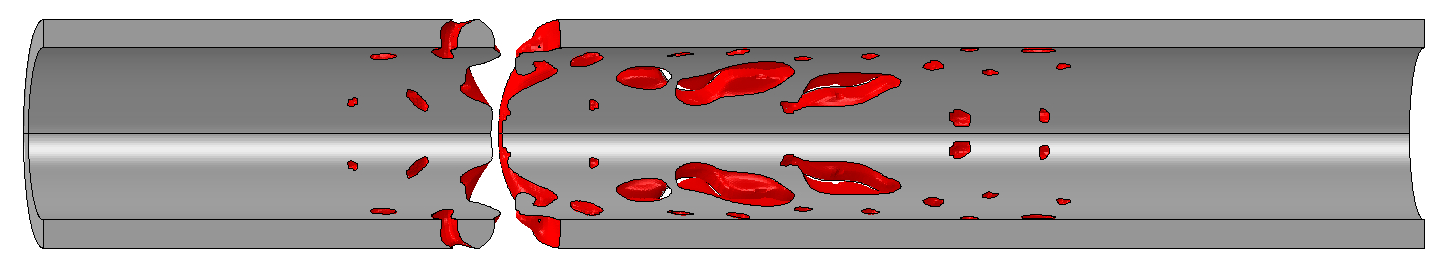}
    \end{center}
 \hspace{0.5cm} (c)
    \end{minipage}\\[0.2cm]
    \begin{minipage}{0.8\linewidth}
    \begin{center}
        \includegraphics[width=\linewidth]{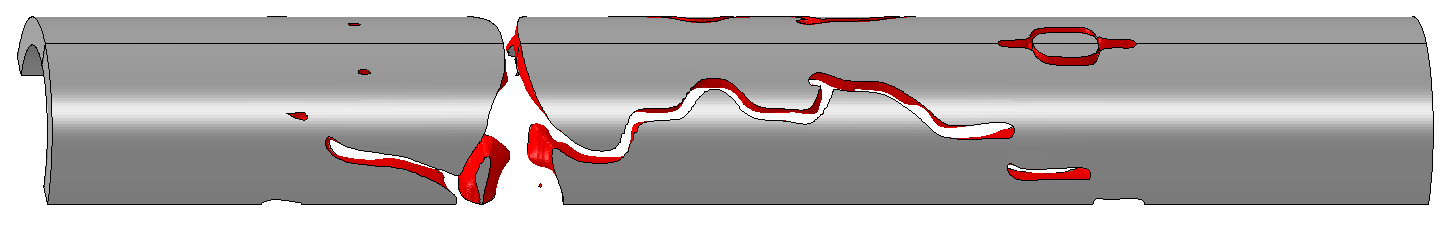}
    \end{center}
 \hspace{0.5cm} (d)
    \end{minipage}
    \caption{Crack initiation and growth predicted in a pipeline with an initial distribution of defects. Cracks are visualized by removing areas with $\phi > 0.8$. Each of the subfigures corresponds to an internal pressure of: (a) 22.95 MPa, (b) 72.25 MPa, (c) 74.8 MPa, and (d) 85 MPa.}
    \label{fig:PipelineHoles}
\end{figure}

As shown in Figs. \ref{fig:Pipeline} and \ref{fig:PipelineHoles}, finite element predictions reveal that damage initiates at a few critical defects where the local stress concentrations increase the local concentration of hydrogen, causing the defects to eventually grow and merge. The coalescence of defects rapidly propagates the damage, leading to a complete failure of the cross-section and the appearance of axial cracks. 

\section{Conclusions}
\label{Sec:Conclusions}

We have presented a generalised phase field formulation for predicting hydrogen assisted fracture. The coupled deformation-diffusion-damage framework developed considers, for the first time, both AT1 and AT2 phase field models, the role of inertia, and a 3D finite element implementation. By addressing four case studies of particular interest, the capabilities of phase field-based models in opening new horizons in structural integrity assessment are showcased. Specifically, we demonstrate that the model can: (i) capture the complex cracking patterns resulting from dynamic loading of an embrittled material, (ii) predict advanced fracture phenomena such as crack kinking, the interaction between neighboring defects and unstable fracture, (iii) conduct virtual experiments involving contact, friction and multiple components, and (iv) simulate in-service conditions, including the current damage state of large scale engineering components. One notable strength of the framework is the possibility of introducing existing defects by assigning an initial value to the phase field variable, without the need of \textit{ad-hoc} and complicated finite element geometries/meshing. This enables a smooth coupling with inspection data and the development of so-called \emph{Digital Twins} of critical structural elements. The results suggest that multi-physics phase field-based simulations can be key in preventing catastrophic failures, enabling virtual fitness-for-service assessment and optimising material selection, structural design and inspection planning.

\section{Acknowledgments}
\label{Sec:Acknowledge of funding}

The authors gratefully acknowledge financial support from the Danish Hydrocarbon Research and Technology Centre (DHRTC). E. Mart\'{\i}nez-Pa\~neda additionally acknowledges financial support from the EPSRC (grants EP/R010161/1 and EP/R017727/1) and from the Royal Commission for the 1851 Exhibition (RF496/2018).



\bibliographystyle{elsarticle-num} 
\bibliography{library}


\end{document}